\documentclass[11pt, letterpaper]{article}
\usepackage{natbib,geometry,rotate,lscape,epsfig,
 colortbl, xspace,sectsty, amsthm,float,xy,xr,
amssymb,amsfonts,amsbsy,amsmath,hyperref,delarray,times,
bm,color,multirow,latexsym,fancyhdr,fontenc, anysize, url, subfigure,
rotating,ifthen,fancybox,fullpage,amscd,pifont}
\usepackage[plain,noend]{algorithm2e}

\newcommand{\no}{\noindent}

\newcommand{\bc}{\begin{center}}
\newcommand{\ec}{\end{center}}

\newcommand{\la}{\label}

\newcommand{\mb}{\boldsymbol}

\newcommand{\be}{\begin{eqnarray}}
\newcommand{\ee}{\end{eqnarray}}

\def\ed{\end{document}}
\def\ci{\cite}
\def\cp{\citep}

\def\bco{\iffalse}

\def\references{\bibliography{f-optdes}}

\def\S{\mathcal{S}}

\def\wh{\widehat}

\def\Var{\hbox{Var}}

\newcommand{\R}{\right}

\def\ss{\subsection*}
\def\sss{\subsubsection*}

\newtheorem*{thm*}{Theorem}

\def\Cov{\hbox{Cov}}
\def\cov{\hbox{cov}}\def\MISE{\hbox{MISE}}
\def\mb{\mathbf}
\def\R{\mathcal{R}}

\bibpunct{(}{)}{;}{a}{}{,}

\makeatletter
\def\singlespace{\def\baselinestretch{1}\@normalsize}

\renewcommand{\baselinestretch}{1.5}

\begin{document}

\thispagestyle{empty}  \bc {\bf \sc \Large Optimal Designs for Longitudinal
and Functional Data}\footnote{Research supported by NSF grants DMS-1104426 and DMS-1407852} \vspace{0.25in}\ec

\vspace*{0.1in} \centerline{April  2016} 
\centerline{Second Revision}\vspace*{0.1in}
\begin{center}
Hao Ji\footnote{Corresponding Author}\\
Department of Statistics\\
University of California, Davis \\ %
One Shields Avenue\\
Davis, CA 95616 U.S.A. \\ %
Phone: 1 (530) 400-5942\\ %
Fax: 1 (530) 752-7099\\
Email: haoji@ucdavis.edu\\

\vspace*{0.3in} Hans-Georg M\"uller\\
Department of Statistics \\ %
University of California, Davis \\ %
One Shields Avenue\\
Davis, CA 95616 U.S.A. \\ %
Phone: 1 (530) 752-2361\\
Fax:  1 (530) 752-7099\\
Email: mueller@wald.ucdavis.edu \\
\vspace*{0.3in}
\end{center}\vspace{.5in}

\thispagestyle{empty}
\newpage \pagenumbering{arabic}

\vspace{0.05in} \thispagestyle{empty} \bc {\bf \sf ABSTRACT} \ec

We propose novel optimal designs for longitudinal data for the common situation where the resources for longitudinal data collection are limited, by determining the optimal locations in time where measurements should be taken.  As for all optimal designs, some prior information is needed to implement the proposed optimal designs. We demonstrate that this prior information may come from a pilot longitudinal study that has irregularly measured and noisy measurements, where for each subject one has available a small random number of repeated measurements that are randomly located on the domain. A second possibility of interest is that  a pilot study consists of densely measured functional data and one intends to take only a few measurements at strategically placed locations in the domain for the future collection of similar data.  We construct optimal designs by targeting two criteria:



(a) Optimal designs to recover the unknown underlying smooth random trajectory for each subject from a few optimally placed  measurements such that squared prediction errors are minimized;

(b) Optimal designs that minimize prediction errors for functional linear regression with functional or longitudinal predictors and scalar responses, again from a few optimally placed measurements.

The proposed optimal designs address the need for sparse data collection when planning longitudinal studies,  by taking advantage of the close  connections between longitudinal and functional data analysis.  We demonstrate in simulations that the proposed designs perform considerably better than randomly chosen design points and
include a motivating data example from the Baltimore longitudinal study of aging. The proposed designs are shown to have an asymptotic optimality property.

\vspace{0.1in}

\no {KEY WORDS:\quad Asymptotics, Coefficient of Determination, Functional Data Analysis, Functional Principal Components, Gaussian Process, Karhunen-Lo\`eve Expansion, Longitudinal Data, Prediction Error, Sparse Design}.\\
\thispagestyle{empty} \vfill

\ss{1. Introduction}

Functional data analysis has become increasingly useful in various fields. In many applications, especially in longitudinal studies, often only a few repeated measurements can be obtained  for each subject or item,  due to cost or logistical constraints that limit the number of measurements.   In some  functional/longitudinal data where the recordings  are sparse and have been taken at irregular time points,  functional data analysis methodology has proved useful to infer covariance structure and trajectories \cp{staniswalis:1998,yao:05:1,li:10}.  While ideally, longitudinal and functional data would be measured on a dense grid, in practical studies one usually encounters constraints on data collection. It is then of interest to have criteria and principles to determine where on the domain
 (usually but not necessarily  a time interval) one should place a given number of measurements so as to minimize prediction errors when recovering  the unobserved trajectories for each subject or to predict a response that is associated with each longitudinal trajectory.  Sparse sampling is also of interest in applications where one has available densely sampled functional data, but can sample at only a few important locations in future longitudinal data collection. The question we address here is where these locations should be. Answering these questions is for example of keen interest to determine optimal monitoring schedules for children in developing countries, aiming to recover their growth trajectories from sparse measurements in order to assess growth stunting and faltering. 

\bco

functional methods based on densely measured functional data are usually not applicable. This research project provides a method of choosing finitely many design points based on sparse functional data, aiming at consistent trajectory recovery and scalar response prediction.\\

A generally useful method is functional principal component analysis for longitudinal data \cp{yao:05:1,yao:05:2}, where trajectory recovery and functional regression problems are addressed by extracting information from the pooled sample to obtain consistent estimates of mean and eigenfunctions of the covariance operator for the population. Local linear smoothing techniques are used to estimate the population mean and covariance functions instead of individual curves to cope with the sparsity in the data. Then functional principal components are targeted via their best linear predictors, conditioning on the available measurements for an individual.

\fi

Several previous studies have discussed methods and algorithms for finding optimal designs in the dense functional data case where data are sampled on a dense regular grid \cp{Ferr:10,hall:12:1}, or belong to a special family of functions \cp{McKea:2010}, where  the emphasis has been on nonparametric functional regression and classification. However, these methods cannot be extended to the case of sparse functional/longitudinal data, which  is a case of crucial interest, as the design selection impacts the planning of longitudinal studies that are often cost-intensive.  A crucial design feature is where to place a limited number of future longitudinal observations. While the connections between longitudinal and nonparametric approaches are being increasingly studied \cp{guo:04:1,xian:13} and there are also studies on designs for classical parametric longitudinal models \cp{ment:97, anis:07} and for random processes and fields \cp{zago:09, fedo:13}, to  our knowledge, there is no previous work on optimal designs for longitudinal studies when the underlying  longitudinal trajectories are viewed as smooth random functions.
The proposed optimal designs fill this gap and are found to perform well in simulations (see Section 5).

Our approach is motivated by the idea of design selection for longitudinal studies such as the well-known Baltimore Longitudinal Study of Aging (BLSA) \cp{Shock:1984, Pearson:1997}, where (among other variables) body mass index profiles are  measured sparsely in time. Currently available data essentially feature random timing of the measurements  (see Figure 1). Our methods will be useful to determine optimal designs for a follow-up study or to recruit new subjects, where we develop optimal designs for (1) recovering the unknown smooth underlying trajectories, as these cannot be easily obtained with nonparametric methods due to the sparseness of the measurements,  and (2)  for predicting scalar responses in functional linear models. As we will demonstrate,  sparsely and irregularly sampled data from a pilot study suffice to construct consistent estimates for the optimal designs. A design resulting from our methodology is depicted in Figure 1, where the three highlighted locations are optimally placed to recover the underlying smooth BMI trajectories when one is constrained to select just  three measurements of BMI in future studies.

\begin{center}
\begin{figure}[H]
\label{BLSA_spa_des}
\centering
\includegraphics[scale=0.25]{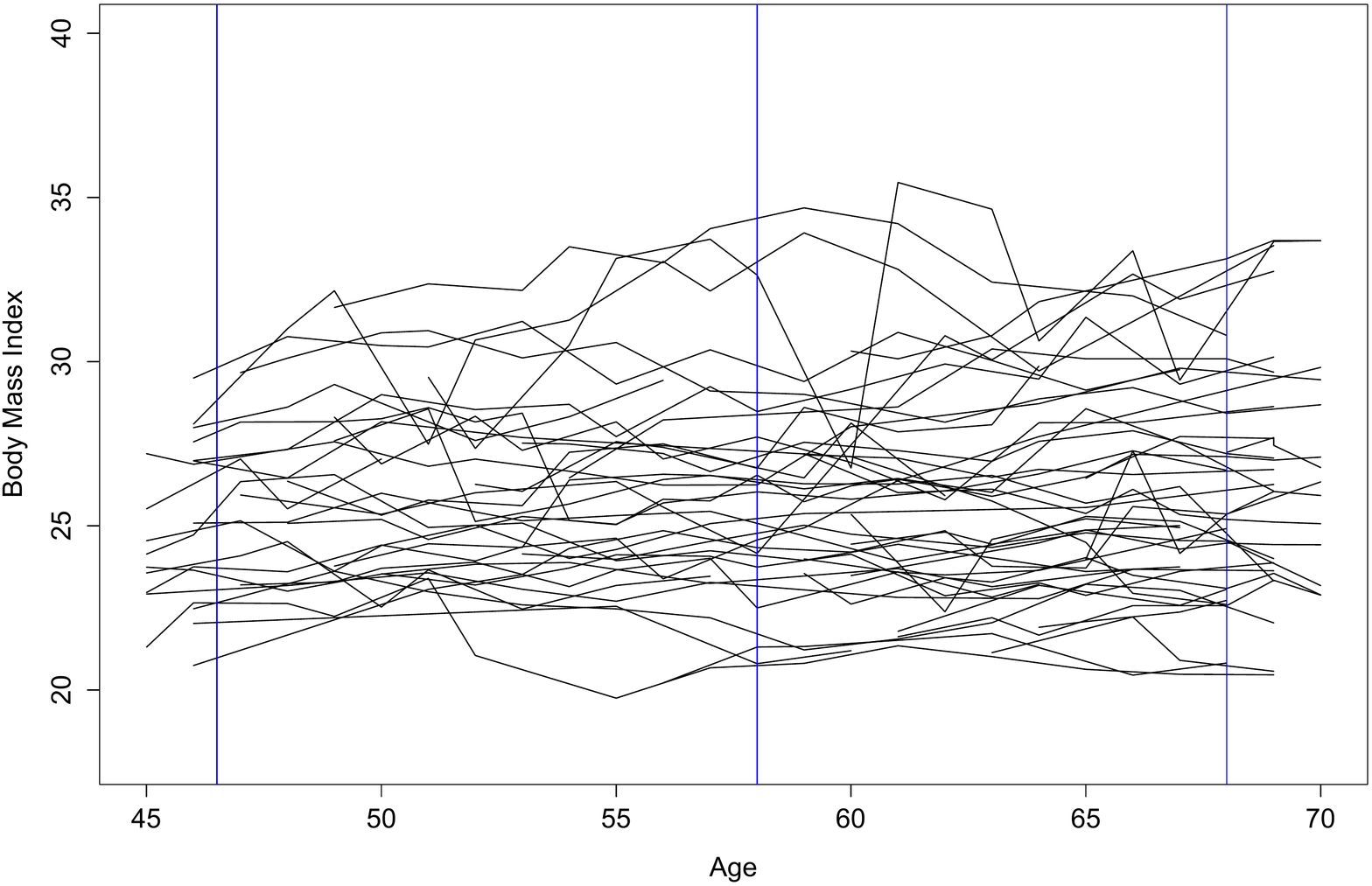}
\caption{Spaghetti plot for a subset of body mass index (BMI)  profiles observed in the Baltimore Longitudinal Study of Aging. The blue vertical bars denote the estimated locations of the optimal design points  for recovering BMI trajectories, among  designs  with three measurement locations.}
\end{figure}
\end{center}

\vspace{-1cm}

\bco

We assume that our sparsely sampled observations are generated by two independent random mechanisms. In a first random selection, a sample of smooth square integrable trajectories is drawn from an underlying random process that has a compact interval $\mathcal{T}$ as its domain. It is important to note that in longitudinal settings we do not observe the trajectories $X_i$. Instead, there is a second random mechanism that selects a random number $m_i$ and random locations $(t_{i1},\ldots,t_{im_i})$ of observations for the $i$-th subject and also independent measurement errors $(e_{i1},\ldots,e_{im_i})$. Here the $m_i$, $X_i$, errors and locations are all independent and i.i.d. Then the $m_i$ repeated observations for the $i$-th subject are $X_i(t_{i1})+e_1,\ldots,X_i(t_{im_i})+e_{m_i}$, where $E(e_{ij})=0$, $\Var(e_{ij})=\sigma^2$, $j=1,2,\ldots,m_i$.

In this paper, we aim at constructing optimal designs for
(1) trajectory recovery of the underlying but unobservable smooth random trajectories, and
(2) linear functional regression with scalar response.
Throughout, we assume that the cost or logistic limitations restrict collection of data to only $p$ measurement per subject and that pilot information is available, either from a sample of densely sampled functional data or from a sample of longitudinal data where the measurements may be sparse and have been made at
random locations, as is the case for the BLSA. All statistical design considerations  require pilot information from a previous study, this is already the case when one simply wishes to determine the sample size needed for a clinical trial, where for example the variances of key outcome variables need to be available from a pilot study.

To address the question  where a given number of $p$ measurements is optimally placed on the time domain 
for specific target criteria, 
we refer to the 

\fi
The vector of function values $\mb{X}=(X(t_1),\ldots,X(t_p))^T$ of a generic process $X$ at design points $\mb{t}=(t_1, \ldots, t_p)^T$, where $^T$ denotes transpose, is not observed in practice, as observations are contaminated with measurement errors. We refer to the observed values at the design points as  $\mb{U}=(U(t_1),\ldots,U(t_p))^T$,  where \be \la{data} U(t_j)=X(t_j)+e_j, \quad Ee_j=0, \, \Var(e_j)=\sigma^2, \, j=1, \ldots, p,\ee
and errors $e_j$ are independent. 
For trajectory recovery, we use best linear predictors that we denote by $B$ \cp{rice:2001} 
\begin{equation} \label{condexp1}
B(X(t)|\mb{U})=\mu(t) + \Cov(X(t),\mb{U})\Cov(\mb{U})^{-1}(\mb{U}-\boldsymbol{\mu}), \quad \mu(t) = {E}(X(t)),\end{equation}
with $\boldsymbol{\mu} = E{\bf{U}} = (\mu(t_1),\mu(t_2),...,\mu(t_p))^T$; for details of the derivation see Online Supplement A.1. Under Gaussian assumptions these are also the best predictors,  as then
$E(X(t)|\mb{U})=B(X(t)|\mb{U}).$
Optimal designs for trajectory recovery are  derived by minimizing the expected squared distance between these best linear predictors and the true trajectories, which turns out to be  equivalent to maximizing a generalized coefficient of determination with respect to the design points (for details see Online Supplement A.3).

For response prediction, where a functional predictor is coupled with a scalar response, the functional linear model is a classical approach \cp{card:99,ramsayfunctional},
\begin{equation}\label{flm}
E(Y|X)=\mu_Y+\int_\mathcal{T} \beta(t)X^c(t)dt,
\end{equation}
where $\mu_Y=E Y$, $\beta(t)$ is the regression coefficient function, and $X^c(t)$ is the centered predictor process, i.e. $X^c(t)=X(t)-\mu(t)$. The difficulty in longitudinal designs is that the integral on the r.h.s of (\ref{flm}) cannot be evaluated, whenever there is sparse sampling of trajectories $X_i$. Therefore, for sparse designs, we condition on $\mb{U}$ on both sides of the model in (\ref{flm}), giving the sparse version 
\begin{equation} \label{condexp}
B\{E(Y|X)|\mb{U}\}=\mu_Y + \int_\mathcal{T} \beta(t)B(X^c(t)|\mb{U}) dt,
\end{equation}
where again, under the Gaussian assumption, $E(X^c(t)|\mb{U}) = B(X^c(t)|\mb{U})$ and $E\{E(Y|X)|\mb{U}\} = B\{E(Y|X)|\mb{U}\}$.
Then the optimal designs $\mb{t}=\{t_1,\ldots, t_p\}$  are defined as the minimizers of the squared prediction error  $E(Y-B\{E(Y|X)|\mb{U}\})^2$. 


The paper is organized as follows. We discuss optimal designs for trajectory recovery and response prediction in Sections 2 and 3, respectively, followed by estimated optimal designs in section 4 and numerical implementations in Section 5, with simulation studies in Section 6.  Data analysis examples from various areas 
are in Section 7, and asymptotic results in Section 8.

\ss{2. Optimal Designs for Trajectory Recovery}
For fixed $p$, consider a generic set of non-random design points $\mb{t}=(t_1,\ldots,t_p)^T$, and corresponding values of the underlying process, $\mb{X} = (X(t_1),\ldots,X(t_p))^T$, with noisy observations $\mb{U}=(U(t_1),\ldots,U(t_p))^T$ and mean vector  $\boldsymbol{\mu}=(\mu(t_1),\ldots,\mu(t_p))^T$.
We aim to find optimal designs $\mb{t}_\mb{X}^*$ with respect to a target criterion. For the underlying process $X$ we write $\mu(t)=EX(t)$ for the mean and $\Gamma(s,t)=\cov(X(s),X(t))$ for the auto-covariance function, 
and the covariance matrices $\boldsymbol{\Gamma}= \hbox{Cov}(\mathbf{X})$ and  
$\boldsymbol{\Gamma}_*=\hbox{Cov}(\mathbf{U})$, where we note that (\ref{data}) implies  $\boldsymbol{\Gamma}_* = \boldsymbol{\Gamma}+\sigma^2 I_p$, with  $I_p$ denoting the $p \times p$ identity matrix. 
With $\mu(t)=EX(t)$, the best linear predictor in (\ref{condexp1}) becomes 
\begin{eqnarray} \nonumber 
B(X(t)|\mb{U})
&=& q(t) + \boldsymbol{\alpha}(t)^T \mb{U}, \quad\quad q(t)=\mu(t)-\boldsymbol{\gamma}(t)^T \boldsymbol{\Gamma}_*^{-1}\boldsymbol{\mu},\\
\label{condcondtraj} \boldsymbol{\alpha}(t)&=&\boldsymbol{\Gamma}_*^{-1}\boldsymbol{\gamma}(t), \quad\quad
\boldsymbol{\gamma}(t) =\Cov(\mb{U},X(t)) =(\Gamma(t_1,t),\ldots,\Gamma(t_p,t))^T,
\end{eqnarray}
where $\boldsymbol{\gamma}(t)$ is a $p$-dimensional vector of covariances associated with the non-random time points $\mb{t}=(t_1,\ldots,t_p)$. Here and in the following, expectations and covariances are considered to be conditional on designs $\mb{t}$.

Our goal is to minimize the mean integrated squared error (MISE) of recovered trajectories that are obtained with the best linear predictors $B(X(t)|\mb{U})$ as a function of the design points $\mb{t}=(t_1,\ldots,t_p)^T$ that exert their influence through $\mb{X}=(X(t_1),\ldots,X(t_p))^T$, i.e. to minimize
\begin{equation} \label{MISE}
\MISE(\mb{t}) = E \int_\mathcal{T} [X(t) - B(X(t)|\mb{U})]^2 dt.
\end{equation}
The performance of recovered trajectories at a fixed point $t \in \mathcal{T}$ can be quantified  by a point-wise coefficient of determination, defined as
\begin{equation}
R^2(t) = \frac{\Var(B(X(t))|\mb{U})}{\Var(X(t))}
\end{equation}

This motivates  to maximize an overall coefficient of determination with respect to the design points. It is easy to see that 
\begin{equation}\label{integR2}
R_X^2= \frac{\int_\mathcal{T} \boldsymbol{\gamma}(t)^T \boldsymbol{\Gamma}_*^{-1} \boldsymbol{\gamma}(t) dt}{\int_\mathcal{T}\Var(X(t)) dt},
\end{equation}
 is a well-defined coefficient of determination for all $\mb{t} \in \mathcal{T}(\delta_0)$ for some $\delta_0 > 0$, where
$\mathcal{T}(\delta_0)$ is defined as
\be \mathcal{T}(\delta_0)=\{(t_1,\ldots,t_p)^T \in \mathcal{T}^p, \,\,  \Cov(U(t_1),\ldots,U(t_p))-\delta_0 I_p \,\, \text{ is positive definite}\}. \la{Td} \ee
Requiring $\mb{t} \in \mathcal{T}(\delta_0)$ ensures uniform matrix invertibility across designs. We assume that the true optimal design also satisfies $\mathbf{t}_X^0 \in \mathcal{T}(\delta_0)$.
In Online Supplement A.3 we show that minimizing the MISE in dependence on $\mb{t}$ as in (\ref{MISE}) is equivalent to maximizing $R_X^2$ in (\ref{integR2}), which is in turn equivalent to find
\begin{equation}\label{optdestraj}
\mb{t}_X^*=(t_{X_1}^*,\ldots,t_{X_p}^*)^T=\hbox{argmax}_{(t_1,\ldots,t_p)^T \in \mathcal{T}(\delta_0)} \int_\mathcal{T} \boldsymbol{\gamma}(t)^T \boldsymbol{\Gamma}_*^{-1} \boldsymbol{\gamma}(t) dt.
\end{equation}

\ss{3. Optimal Designs for Predicting Scalar Responses}
We assume here the same setting and conditions as in the previous section and aim at finding optimal designs $\mb{t}_Y^*$ with respect to a target criterion that is specific for the functional linear model in  (\ref{flm}).

The best linear predictor in (\ref{condexp}), using  (\ref{condcondtraj}), is seen to be 
\begin{eqnarray}\label{lmpred}
B\{E(Y|X)|\mb{U}\} &=& \mu_Y + \int_\mathcal{T} \beta(t)B(X^c(t)|\mb{U})dt \nonumber \\
    &=& \mu_Y + \int_\mathcal{T} \beta(t)\boldsymbol{\alpha}(t)^T (\mb{U}-\boldsymbol{\mu}) dt
    = \mu_Y + \boldsymbol{\beta}{}_p^T (\mb{U}-\boldsymbol{\mu}),\end{eqnarray}
with 
$\boldsymbol{\beta}{}_p = \int_\mathcal{T} \beta(t)\boldsymbol{\alpha}(t) dt.$
A basic tool for the following derivations is the Karhunen-Lo\`eve expansion of square integrable random processes,
\begin{equation}\label{KLX}
X(t) = \mu(t) + \sum_{k=1}^{\infty} \zeta_k \psi_k(t),
\end{equation}
with eigenfunctions $\psi_k$, $k=1,2,\ldots$ of the covariance operator of $X$, and uncorrelated (independent in the Gaussian case) functional principal components (FPCs) $\zeta_k, \, k=1,2,\ldots$. The eigenfunctions form an orthonormal basis of the space $\mathcal{L}^2(\mathcal{T})$ and one has the covariance expansion
\begin{equation} \label{AutoCOV}
\Gamma(s,t) = \Cov(X(s),X(t))= \sum_{k=1}^{\infty} \rho_k \psi_k(s) \psi_k(t),
\end{equation}
where the eigenvalues $\rho_k$ of the covariance operator are positive and ordered, $\rho_1 > \rho_2 > \ldots$. The FPCs satisfy $E(\zeta_k)=0$ and $\Var(\zeta_k)=\rho_k$ for all $k$.

The regression coefficient function $\beta(t)$ of the FLM (\ref{flm}) can be expanded in the eigenbasis representation  \cp{he:2000}, with convergence under mild regularity conditions, 
\begin{equation}\label{betaexp}
\beta(t) = \sum_{k=1}^{\infty} \frac{E(\zeta_kY)}{E(\zeta_k^2)} \psi_k(t).
\end{equation}
Observing that for the cross-covariance function
\begin{equation}
C(t) = \Cov(X(t), Y)=\Cov(\sum_{k=1}^{\infty} \zeta_k\psi_k(t), Y)=\sum_{k=1}^{\infty}E(\zeta_kY)\psi_k(t)
\end{equation}
and using the orthonormality of the eigenfunctions $\psi_k$, it is easy to see that $\sigma_k = E(\zeta_kY)$ can be written as
\begin{equation}\label{sigmak}
\sigma_k = \int_\mathcal{T} C(t)\psi_k(t) dt.
\end{equation}

By (\ref{condcondtraj}),   (\ref{betaexp}) and (\ref{sigmak}), we have with  $\boldsymbol{\psi}_k = (\psi_k(t_1),...,\psi_k(t_p))^T$ and $\mathbf{C} = (C(t_1),...,C(t_p))^T$, \begin{eqnarray}\label{betap}
\boldsymbol{\beta}_p&=&\int_\mathcal{T} \beta(t)\boldsymbol{\alpha}(t)dt
    = \boldsymbol{\Gamma}_*^{-1} \int_\mathcal{T} (\sum_{k=1}^{\infty} \frac{\sigma_k}{\rho_k}\psi_k(t)) (\sum_{k=1}^{\infty} \rho_k \psi_k(t) \boldsymbol{\psi}_k) dt \nonumber \\
    &=& \boldsymbol{\Gamma}_*^{-1} \sum_{k=1}^{\infty} \sigma_k \boldsymbol{\psi}_k
    = \boldsymbol{\Gamma}_*^{-1} \mb{C}.
\end{eqnarray}
Similar to trajectory recovery in the last section, we propose to minimize the prediction error
\begin{equation} \label{prederror}
E [ Y-B\{ E(Y|X)|\mb{U} \} ]^2 = E [ Y- \mu_Y - \boldsymbol{\beta}{}_p^T(\mb{U}-\boldsymbol{\mu}_\mb{X}) ]^2. 
\end{equation}
This is shown in Online Supplement A.3 to be equivalent to maximizing the following coefficient of determination $R_Y^2$ that quantifies prediction power,
\begin{equation}\label{R2Y}
R_Y^2 = \frac{{\Var[B\{E(Y|X)|\mb{U}\}]}}{{\Var(Y)}},
\end{equation}
where we  assume that the true optimal designs for regression case $\mathbf{t}_Y^0$ lies in $\mathcal{T}(\delta_0)$.

To maximize $R_Y^2$, it is equivalent to find
\begin{eqnarray}\label{optdespred}
{\mb{t}}_Y^* = \hbox{argmax}_{(t_1,\ldots,t_p)^T \in \mathcal{T}(\delta_0)} {\Var[B\{E(Y|X)|\mb{U}\}]} \nonumber 
         = \hbox{argmax}_{(t_1,\ldots,t_p)^T \in \mathcal{T}(\delta_0)} \boldsymbol{\beta}{}_p^T \boldsymbol{\Gamma}_* \boldsymbol{\beta}_p.
\end{eqnarray}

\noindent Therefore, by (\ref{betap}), the optimization criterion can be simplified to
\begin{equation}\label{optpred}
\mb{t}_Y^* = \hbox{argmax}_{(t_1,\ldots,t_p)^T \in \mathcal{T}(\delta_0)} \mb{C}^T \boldsymbol{\Gamma}_*^{-1} \mb{C}.
\end{equation}

\ss{4. Estimated Optimal Designs}
While the population optimal designs were derived in the previous sections, in practice they must be estimated from available data. The available observations are 
\begin{eqnarray}\label{obs}
U_{ij} &=& X_i(t_{ij}) + e_{ij}, \,\, 1 \leq i \leq n, \,\, 1 \leq j \leq m_i, \\
Y_{i} &=&\mu_Y+\int_\mathcal{T} \beta(t)X_i^c(t)dt + \epsilon_i, \,\, 1 \leq i\leq n, \label{obsY}
\end{eqnarray}
where (\ref{obsY}) only applies to the prediction scenario. Here $(X_i, Y_i), \,\, i=1,\ldots,n,$ are independent realizations of $(X, Y)$, with $m_i$ the number of observed function values for each subject, the $t_{ij}$ are randomly located time points on $\mathcal{T}$ with density function $f_T(\cdot)$, and the $e_{ij}$ and the $\epsilon_{ij}$ are random errors with zero mean and variance $\sigma^2$ and $\sigma_Y^2$, respectively. We assume throughout that the $(X_i, Y_i)$, the $e_{ij}$ and the $\epsilon_{i}$ are all independent. A notable feature of this data model is that it includes noise not only in the responses $Y_i$ but also in the recordings of the random trajectories. 

From the data, estimates of mean function $\mu(t)$, auto-covariance function $\Gamma(s,t)$ and cross-covariance function $C(t)$ are obtained on a user-defined fine grid covering $\mathcal{T}$ and  these are denoted as $\widehat{\mu}(t)$, $\widehat{\Gamma}(s,t)$ and $\widehat{C}(t)$, respectively. Consistent estimates of these quantities from the pilot study are needed to obtain consistent estimates of the  optimal designs. For densely observed functional data, cross-sectional estimates are suffcient. Methods to overcome the difficulty of sparse sampling when targeting the mean and covariance functions have been addressed by various authors \cp{yao:05:1,yao:05:2,staniswalis:1998,li:10}. The sparsely sampled  case is different from the more commonly considered situation of densely sampled functional data, where individual curves can be consistently estimated by direct smoothing \cp{rice2004functional} and the covariance function is readily estimated by cross-sectional averaging \cp{ramsayfunctional}. 

In the sparse case, these direct approaches do not lead to consistent estimates,  due to the sparseness and lack of balance of the measurements. The way forward is to pool data for estimating mean and covariance functions, borrowing strength from the entire sample. For the required smoothing steps, we adopt one- and two-dimensional local linear smoothing, with further details in the following section.  The estimated auto-covariance function is further regularized by retaining only the positive eigenvalues and eigenvectors of the smoothed covariance function, so that $\widehat{\Gamma}(s,t)$ is non-negative definite. An estimate $\widehat{\sigma}^2$ of $\sigma^2$ is also needed and 
 this is discussed in Section 5. 
 For any $p$-dimensional vector $(t_1,...,t_p)^T$ of design points picked from the user-specified dense grid, we then have estimates $\widehat{\boldsymbol{\mu}}_X$, $\widehat{\boldsymbol{\Gamma}}_*$, $\widehat{\boldsymbol{\gamma}}$ and $\widehat{\mathbf{C}}$ for $\boldsymbol{\mu}$, $\boldsymbol{\Gamma}_*$, $\boldsymbol{\gamma}$ and $\mathbf{C}$, respectively. Then the estimated optimal designs are:
\begin{enumerate}
\item For trajectory recovery:
\begin{equation}\label{estoptdes_traj}
\widehat{\mathbf{t}}_X^* = \hbox{argmax}_{(t_1,...,t_p)^T} \int_{\mathcal{T}} \widehat{\boldsymbol{\gamma}}^T(t) \widehat{\boldsymbol{\Gamma}}_*^{-1} \widehat{\boldsymbol{\gamma}}(t) dt,
\end{equation}
where all integrals are implemented with trapezoidal integration.
\item For scalar response regression:
\begin{equation}\label{estoptdes_reg}
\widehat{\mathbf{t}}_Y^* = \hbox{argmax}_{(t_1,...,t_p)^T} \widehat{\mathbf{C}}^T\widehat{\boldsymbol{\Gamma}}_*^{-1} \widehat{\mathbf{C}}.
\end{equation}
\end{enumerate}

\ss{5. Numerical Implementation}
\bco Method used for estimating covariances is addressed here, together with two other main numerical implementation issues are discussed for the proposed method: ridge regression for stable covariance matrix inversion and multidimensional optimization.\fi
\sss{5.1 Mean and Covariance Estimation via Smoothing}
First pooling sparse longitudinal data across subjects, we  apply local linear estimators \cp{li:10} to the resulting scatterplots, which depend on a bandwidth $h$ as a tuning (smoothing) parameter. 
Writing $S_p(t, (Q_j,V_j)_{j = 1,...,m}, h)$ for a local linear $q$-dimensional smoother (with $q=1$ or $q=2$) with output at the predictor level $t$ and employing bandwidth $h$ to smooth the scatterplot   $(Q_j,V_j)$, where $Q_j \in \R^q$, we obtain  estimates $\hat{\mu}(t)$ for the mean function $\mu(t)$ as $S_1(t, (t_{ij}, U_{ij})_{i=1,...,n, j = 1,...,m_i}, h_\mu), \ \ t \in \mathcal{T},$  smoothing estimates $\tilde{\Gamma}(s,t)$ for the autocovariance function $\Gamma(s,t)$ as 
$S_2((s,t), (t_{ij}, t_{ik}, U_{ij}U_{ik})_{i=1,...,n, j,k = 1,...,m_i, j \neq k}, h_R) - \widehat{\mu}(s)\widehat{\mu}(t), \ \  s, t \in \mathcal{T},$ and estimates $\hat{C}(t)$ for the cross-covariance function $C(t)$ as
$S_1(t, (t_{ij}, Y_i U_{ij})_{i=1,...,n, j = 1,...,m_i}, h_S) - \widehat{\mu}(t)\widehat{\mu}_Y, \ \ t \in \mathcal{T}.$
We also obtain the estimate  $\widehat{\mu}_Y$ as the  sample mean of the scalar  responses. 

Bandwidths for all smoothing steps are selected by cross validation or generalized cross validation. Details on the smoothing steps are in Online Supplement A.2, and assumptions for establishing  consistency of the above smoothing steps in the longitudinal data context  are provided  in Online Supplement A.4. We used the function FPCA \cp{yao:05:1} in PACE, a free Matlab package (http://www.stat.ucdavis.edu/PACE/) for smoothing and estimating the model components. 

From the estimates  $\tilde{\Gamma}(s,t)$ we then  obtain estimates $\hat{\rho}_j$ and  $\hat{\psi}_j$ for eigenvalues and eigenfunctions of predictor processes $X$ by discretization and matrix spectral decomposition. The final  autocovariance estimates  $\widehat{\Gamma}(s,t)$ are obtained by projecting on the space 
of non-negative and symmetric surfaces, simply by retaining only  the positive eigenvalues and their corresponding eigenvectors \cp{hall:08}, yielding
$\widehat{{\Gamma}}(s,t)=\sum_{j=1,\,  \widehat{\rho}_j>0}^K \widehat{\rho}_j \widehat{\psi}_j(s) \widehat{\psi}_j(t).$

\sss{5.2 Stable Covariance Matrix Inversion}
As a practical implementation  of the matrix inversion condition (\ref{Td}), 
we apply ridge regression \cp{Hoerl:1970}, enhancing the diagonal of the  autocovariance surface $\wh{\Gamma}_*(s,t)$, as  the matrices that need to be inverted are submatrices of this surface. Adding a suitable ridge parameter $\wh{\sigma}_{new}^2$ at the diagonal ensures positive definiteness of all relevant  $p$ by $p$ submatrices, 
\begin{equation}\label{RidgeCov}
\wh{\Gamma}_{*}(s,t) = \wh{\Gamma}(s,t)+\sigma_{\rm new}^2 \delta_{st}.
\end{equation}
Here $\delta_{st}=1$ if and only if $s=t$. The optimization procedures in (\ref{estoptdes_traj}) and (\ref{estoptdes_reg}) are then implemented with $\wh{\Gamma}_{*}(s,t)$ or $\wh{\boldsymbol{\Gamma}}_{*}=\wh{\boldsymbol{\Gamma}}+\sigma_{\rm new}^2 I_p$.
We explored   two options to select the ridge parameter $\sigma_{\rm new}^2$:\vspace{.3cm}

{\bf 1.  Cross-validation.}  For {\it trajectory recovery}, target criteria are average root mean squared error, ARE, and relative average root mean squared error, ARE$^*$, defined as  
\begin{eqnarray} \nonumber
ARE&=&\frac{1}{n} \sum_{i=1}^n \left\{\frac{1}{m_i}\sum_{j=1}^{m_i} (U_i(t_{ij})-\wh{B}_{-i}(X_i(t_{ij})|\mb{U}))^2\right\}^{1/2},\\
ARE^*&=& \sum_{i=1}^n \left\{\frac{1}{m_i}\sum_{j=1}^{m_i} (U_i(t_{ij})-\wh{B}_{-i}(X_i(t_{ij})|\mb{U}))^2\right\}^{1/2} / \sum_{i=1}^n \left\{\frac{1}{m_i}\sum_{j=1}^{m_i} U_i(t_{ij})^2\right\}^{1/2} \label{ARE}\end{eqnarray}
where $(U_i(t_{i1}),\ldots,U_i(t_{im_i}))^T$ and $(\wh{B}_{-i}(X_i(t_{i1})|\mb{U}),\ldots,\wh{B}_{-i}(X_i(t_{im_i})|\mb{U}))^T$ are observed measurements and plugged-in estimated best linear predictors for recovered processes with design $\mb{t}$ at the same time points.
Leave-one-out versions of ARE and ARE$^*$ are easily obtained if one has  densely measured functional data in the pilot study, but are usually not  viable when the pilot study
consists of longitudinal data with sparse measurements, where observed subjects generally will not have recorded measurements at the selected optimal design points.

For {\it scalar response prediction}, average prediction error (APE) and relative APE are natural criteria  that can be easily cross-validated, irrespective of the design of the pilot study,
\begin{eqnarray} \nonumber
APE &=&\left\{\frac{1}{n}\sum_{i=1}^n(Y_i-\wh{B}_{-i}\{E(Y_i|X_i)|\mathbf{U}_i\})^2\right\}^{1/2},\\
APE^* &=&\left\{\sum_{i=1}^n(Y_i-\wh{B}_{-i}\{E(Y_i|X_i)|\mathbf{U}_i\})^2\right\}^{1/2} / \left\{\sum_{i=1}^n Y_i^2 \right\}^{1/2} \label{APE}\end{eqnarray}
where $\wh{B}_{-i}\{E(Y_i|X_i)|\mathbf{U}_i\}$ is the estimated $i$-th response obtained with estimated optimal designs that are obtained   from a training sample leaving out the $i$-th observation.\vspace{.3cm}

 {\bf 2.  Modified Cross-Validation.}  Direct  cross-validation approach is not feasible for the case of sparsely sampled pilot studies. In a modified approach, for each ridge parameter in a candidate set $\Omega$, we repeatedly and randomly partition the training sample $\S$ from the sparsely sampled pilot study into two sets, $\S_A$ and $\S_B$, estimate model components from $\S_A$, and then  find the relatively best design $\mb{t}_{X,\S_B}^*$ or $\mb{t}_{Y,\S_B}^*$ by maximizing the  criteria in (\ref{optdestraj}) or (\ref{optdespred}) over $\mathcal{T}_B$, which is the set of all available designs determined by the random configurations of the  design points as observed for the subjects in the sample $\S_B$. We then recover the trajectory or estimate the response for those subjects in $\S_B$ where there is a match of  the selected design  $\mb{t}_{X,\S_B}^*$ or $\mb{t}_{Y,\S_B}^*$ with the design for that subject.
  
Combining  $L$ different random partitions, mean ARE or APE are used to evaluate the performance of  the ridge parameter choice in  (\ref{RidgeCov}), yielding the selected parameter 
\be \nonumber
\sigma^2_{\rm{new}}=\mathrm{argmin}_{\Omega} \sum_{l=1}^L \left(\frac{1}{n_B}\sum_{i_B=1}^{n_B} \left\{\frac{1}{m_{i_B}}\sum_{j=1}^{m_{i_B}} (U_{i_B}(t_{i_B,j}) -\wh{B}_{\S_A}(X_{i_B}(t_{i_B,j})|\mb{U}_{i_B}))^2\right\}^{1/2}\right),\ee
for trajectory prediction, and 
\be \nonumber
   \sigma^2_{\rm{new}}=\mathrm{argmin}_{\Omega} \sum_{l=1}^L \left(\frac{1}{n_B}\sum_{i_B=1}^{n_B}(Y_{i_B}-\wh{B}_{\S_A}\{E(Y_{i_B}|X_{i_B})|\mathbf{U}_{i_B}\})^2\right)^{1/2}.
\ee  for response prediction. 
Here, $i_B$ is the index of the subjects in $\S_B$ with available measurements at the selected optimal designs $\mb{t}_{X,\S_B}^*$ and $\mb{t}_{Y,\S_B}^*$ for trajectory recovery and prediction, respectively, with $n_B$ denoting the number of such subjects, and $m_{i_B}$ is the number of measurements for the subject with index $i_B$.  The  estimator $\wh{B}_{\S_A}$ is fitted based on data from sample $\S_A$ only, and index sets  $\S_A$ and $\S_B$ depend on the random partition $l$. 
This method worked well in simulations and applications  with longitudinal pilot studies.

\sss{5.3 Sequential Selection of Design Points}
Computationally, once the number of design points $p$ has been specified, both trajectory recovery and scalar response prediction involve $p$-dimensional optimization. Exhaustive search over all combination of grid points is very time consuming when employing optimization algorithms such as simulated annealing \citep{Kirk:1983}.
A faster alternative  is sequential selection, which  is a greedy algorithm, where one searches for a global optimal designs when $p=2$ as an initial step and then adds design points one-by-one iteratively, until the number of design points reaches the desired number. At each step, the target design point is the one that maximizes the selection criteria when adding it to the currently selected design points, which are carried forward unaltered. The sequential method is fast but does not guarantee finding the optimal solution. The performance differences of sequential and exhaustive search  selection were found to be relatively small in simulation studies.  

\ss{6. Simulation Studies}
We study the performance of optimal designs for trajectory recovery and  scalar response prediction under two separate scenarios. In scenario 1 we consider the case where the pilot study generates  densely observed functional data, and in scenario 2 the case where it  generates  sparse longitudinal data. Random trajectories are generated as $X_i(t) = \mu(t)+\sum_{k=1}^{K} \zeta_{ik} \psi_k(t)$, and observed data as  $U_i(t_{ij}) = X_i(t_{ij}) + e_{ij}$, where $e_{ij} \sim N(0,0.25)$.

For the regression case, the response is chosen as $Y_i = \int \beta(t)X_i(t)dt  + \epsilon_i = \zeta_{i,1}-2\zeta_{i,2}+\zeta_{i,3}-2\zeta_{i,4} + \epsilon_i$, i.e., as a linear combination of functional principal components of the process, where $\epsilon_i \sim N(0,0.25)$ are i.i.d. We specify $\mathcal{T} = [0,10]$, mean function $\mu(t)=\frac{1}{2}t^2 + 2\sin t + 3\cos 2t$, $t \in \mathcal T$, and include 10 functional principal components, with eigenvalues $(30,20,12,8,\frac{30}{25},\frac{30}{36},\frac{30}{49},\frac{30}{64},\frac{30}{81},\frac{30}{100})^T$, and corresponding eigenfunctions $\psi_k(t)=\sqrt{\frac{2}{10}}\cos \{(\frac{k}{10})\pi t\}$, for $k=1,2,\ldots,10$ and  generate data for $100$ subjects in the training sample and $1000$ in the testing sample. The measurement locations $t_{ij}$  are assumed to form a dense grid for simulation scenario 1, and are sparse, with a  random number of $4$ to $8$ measurement locations per subject for scenario 2. Figure 9 in Online Supplement A.9 shows the Spaghetti plot of the data for the subjects in one training sample. 

For both trajectory recovery and prediction for functional linear regression, we applied the proposed procedures for the  subjects in the training sample to construct the optimal designs for $p=2,3,4$ with exhaustive search and $p=2,3,...,8$  with sequential search. Each simulation scenario was repeated 100 times. The optimal ridge parameter $\sigma^2_{\rm{new}}$ in (\ref{RidgeCov}) was determined by cross-validation for scenario 1 and modified cross-validation for scenario 2  (see subsection 5.2).  We compare the median performance with regard to (relative) Average Root Squared Error (ARE) and (relative) Average Prediction Error (APE) defined in (\ref{ARE}) and (\ref{APE}) for optimal designs and random designs. These random designs use the same number of design points as the estimated optimal designs, however the locations are sampled from a uniform distribution over all possible locations (we provide further comments on the rationale of comparing with random designs in Online Supplement A.5). The results are summarized in Table 1, and are illustrated in Figure 2 (and also Figure 10 in Online Supplement A.9) for the case where the pilot study is longitudinal with sparsely sampled functional data.


These simulations demonstrate that the proposed optimal designs exhibit better performance than random designs for both trajectory recovery and scalar response prediction, especially for sparse functional data. For trajectory recovery, Figure 10  (in Online Supplement A.9) for sparse pilot designs indicates that recovered trajectories obtained from optimal designs are closer to the underlying true curves than those obtained from median-performance random designs. For scalar response prediction, Figure 2 corroborates the results from Table 1, namely that optimal designs outperform median-performance random designs for the prediction of a subject's response from a few observed measurements only.

The boxplots of ARE and APE for  100 simulation runs in Figure 3 and Figure 10 (in Online Supplement A.9) visualize  the variation of performance over the simulations, comparing optimal and  median performance random designs, and also show the improvement in performance as the number of design points $p$ increases.  For densely observed functional data, there are various penalization schemes possible to select  $p$, minimizing   the sum of ARE (APE) and a penalty that increases with increasing $p$.   However, such schemes are  not directly applicable for longitudinal data, because subjects rarely will have been observed at the  selected design points for trajectory recovery or prediction. In practice, an upper bound for $p$ frequently will be dictated by cost.
Simulation results for the  effect of ridge parameter selection on the performance of optimal designs showed that the proposed selection works well  (see Online Supplement A.7).
\begin{table}[H]
\caption{Comparing Optimal with Random Designs in 100 Simulations, in terms of mean ARE  and relative ARE${}^*$  (\ref{ARE})  (in brackets) for Trajectory Recovery and APE and relative APE${}^*$ (\ref{APE}) (in brackets) for Response Prediction in a Functional Linear Model. Exhaustive Search used.}
\begin{center}
\label{simtable}
\begin{tabular}{|cc|cc|cc|}
  \hline
   Number of Design Points &  & \multicolumn{2}{c}{Mean ARE (ARE${}^*$)} \vline & \multicolumn{2}{c}{Mean APE (APE${}^*$)} \vline \\
  \hline
   & Design   &   Dense   &   Sparse  & Dense & Sparse\\
   \hline
  $p=2$ & Optimal & $1.74(.091)$ & $1.84(.102)$ & $3.37(.272)$ & $9.41(.759)$\\
   & Random & 1.91(.101) & $2.12(.117)$ & 10.65(.870)  & $10.79(.906)$\\
   \hline
  $p=3$ & Optimal & $1.37(.072)$ & $1.59(.088)$ & $2.79(.224)$ & $7.63(.613)$\\
   & Random & 1.65(.087)  & $1.88(.104)$ & 8.56(.687) & $9.80(.793)$\\
   \hline
  $p=4$ & Optimal & $1.02(.054)$ & $1.34(.074)$ & $2.54(.204)$ & $6.87(.553)$\\
   & Random & 1.46(.077) & $1.74(.096)$ & 6.44(.499) & $7.65(.616)$\\
  \hline
\end{tabular}
\end{center}
\end{table}
\vspace{-.5cm}


To summarize the simulation results, optimal designs performed very well and in any case better than random  designs for both trajectory recovery and scalar response prediction. The costs incurred when adopting the much faster sequential search algorithm care quite small. Unsurprisingly,  performance of the optimal designs was seen to improve  with increasing number of design points.

\bco
\fi

\ss{7. Data Illustrations}
\sss{7.1 Mediterranean Fruit Fly Egg-Laying}
The Mediterranean fruit fly data, described in \cp{Carey:2002}, consist of  egg-laying profiles for $1000$ female Mediterranean fruit flies. For each fly, daily measurements on the number of eggs laid during the day are available from birth to death. A biologically relevant regression problem is to utilize the partial egg-laying profile from day $1$ to day $30$  to predict the number of eggs that will be laid during the remaining lifetime for each fly. This yields information about the reproductive potential of the fly at age 30 days, which is related to its evolutionary fitness \cp{kouloussis2011seasonal}.

\begin{center}
\begin{figure}[H]
\label{simSR3sparse}
\centering
\includegraphics[scale=0.25]{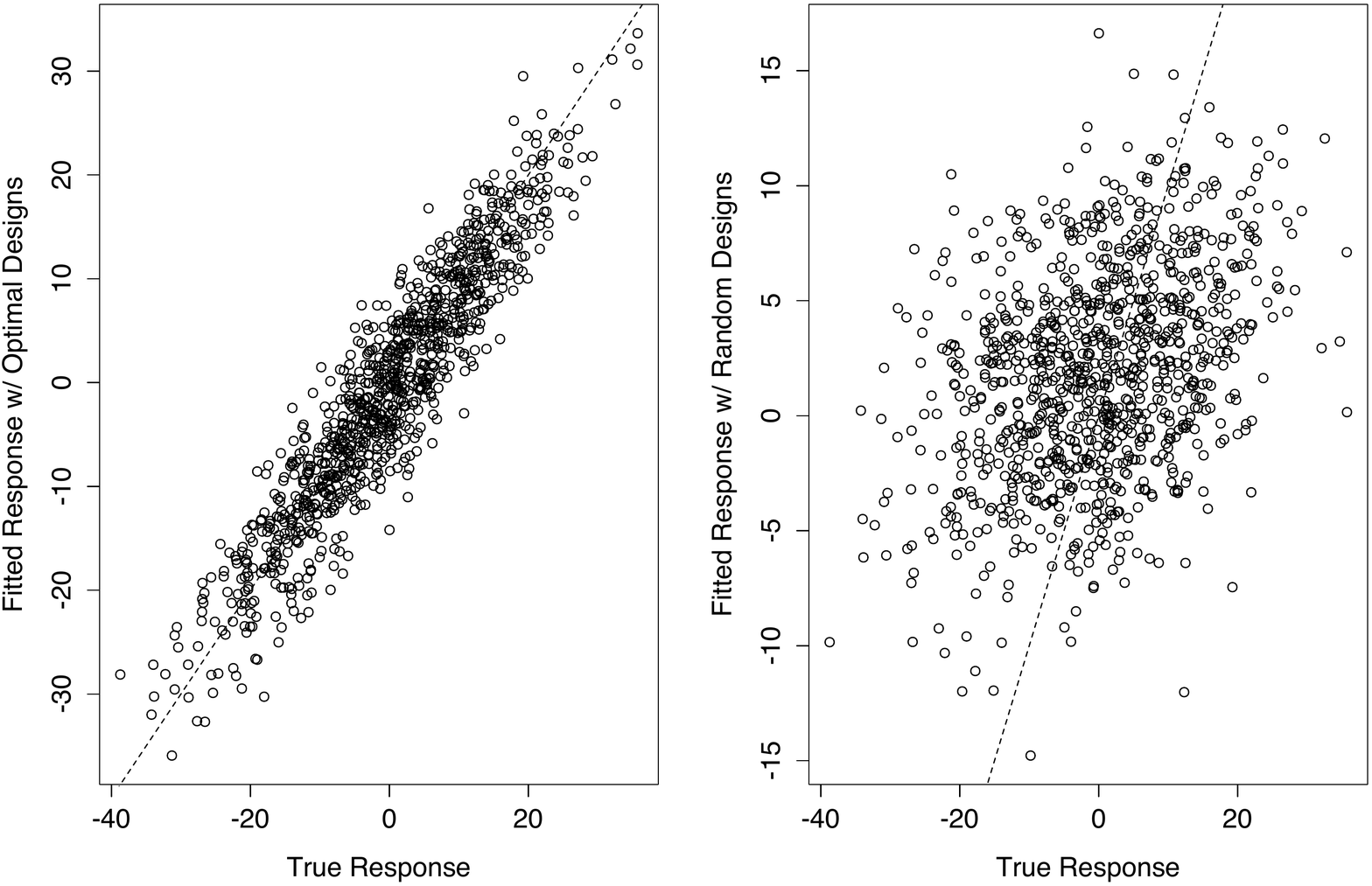}
\includegraphics[scale=0.25]{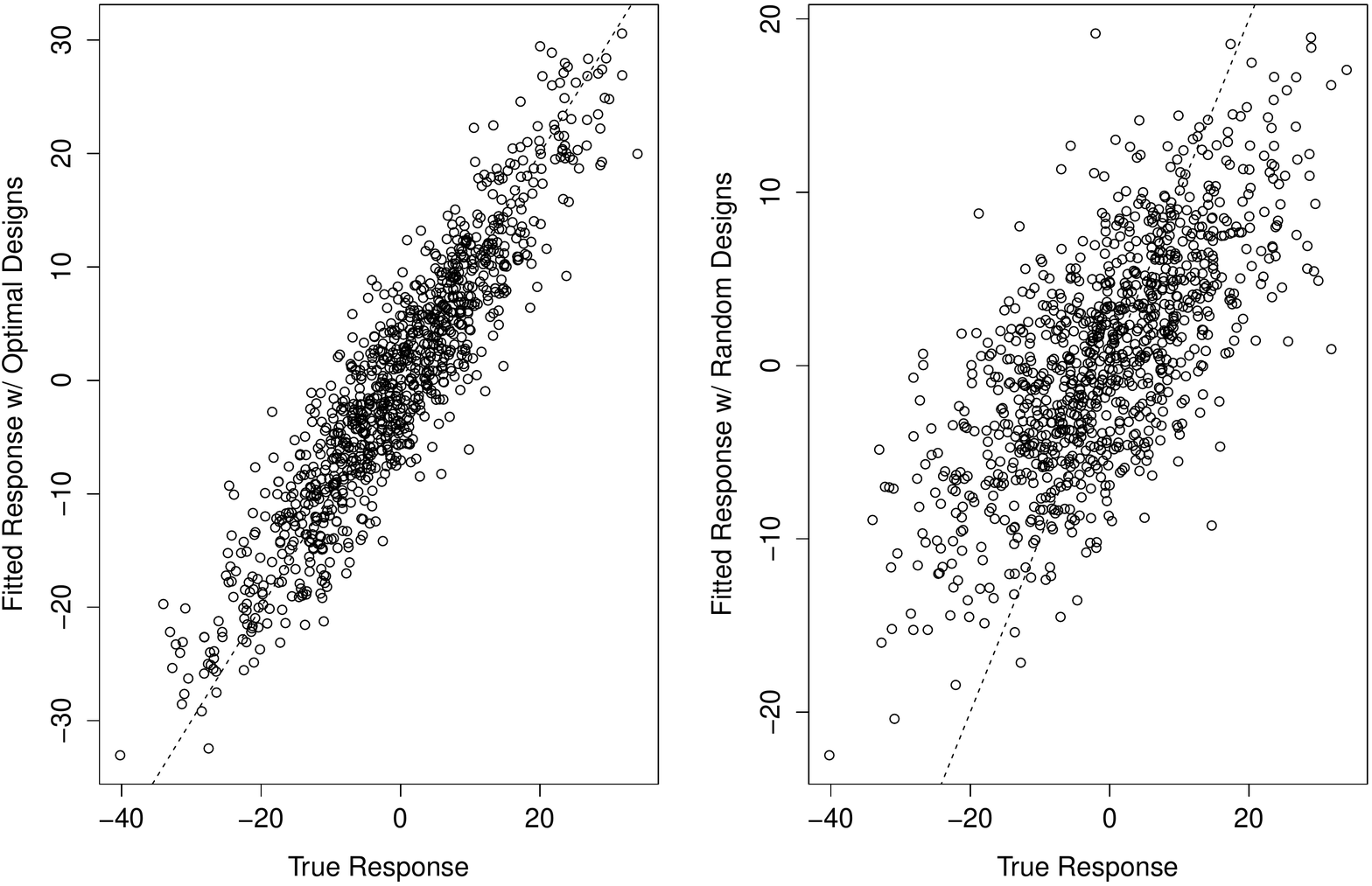}
\caption{Simulation results for sparsely sampled pilot data. Left two (right two) scatterplots are for fitted versus true responses for predicting a scalar response based on 3 (4) design points. First and third panels (from the left) demonstrate the results for optimal designs, while second and fourth panels (from the left) depict the results for random designs with median performance.}
\end{figure}
\end{center}

\vspace{-1.5cm}

We aim to find the optimal design points for this scalar response regression problem. To prevent censoring, we only include flies that live beyond  $30$ days. The measurements in the pilot data are dense and regular. Since daily egg-laying counts require constant monitoring of the flies, reducing this task to monitoring of the flies at a few time points is useful to scale up such studies. It is of additional interest to identify key days that are relevant for the prediction of the egg-laying potential.  We use the complete available egg-laying profiles to find optimal design points that are
most relevant for the prediction of the remaining total number of eggs  via a functional linear regression model.  

From the 667 subjects surviving more than 30 days, we select a training sample of $500$ flies, and a testing sample that consists of the  remaining $167$ flies. Figure 4 shows the Spaghetti plot for a subset of the training sample. We apply the proposed methodology to find optimal designs for $p=3$. 
The relationship between observed and predicted responses is shown in Figure 5.
The relative APE  (\ref{APE}) is $0.483$ when using optimal designs, as opposed to $0.665$ using random designs with median performance for $p=3.$
Figure 5 provides a graphical illustration that optimal designs clearly outperform randomly chosen designs with median performance. The three selected optimal design points for $p=3$ are at days 10, 25 and 26. Their locations are shown as blue vertical lines in Figure 4. These design point locations are consistent with previous findings that both the intensity of egg-laying at earlier ages and the rate of decline at older ages are closely related to the reproductive potential for individual flies \cp{mull:01:3}. Two design points are selected on consecutive days 25 and 26, indicating that these locations are useful to gauge the rate of decline in egg-laying, which corresponds to quantifying a derivative in this age range.

\begin{center}
\begin{figure}[H]\label{simbox}
\centering
\includegraphics[scale=0.2]{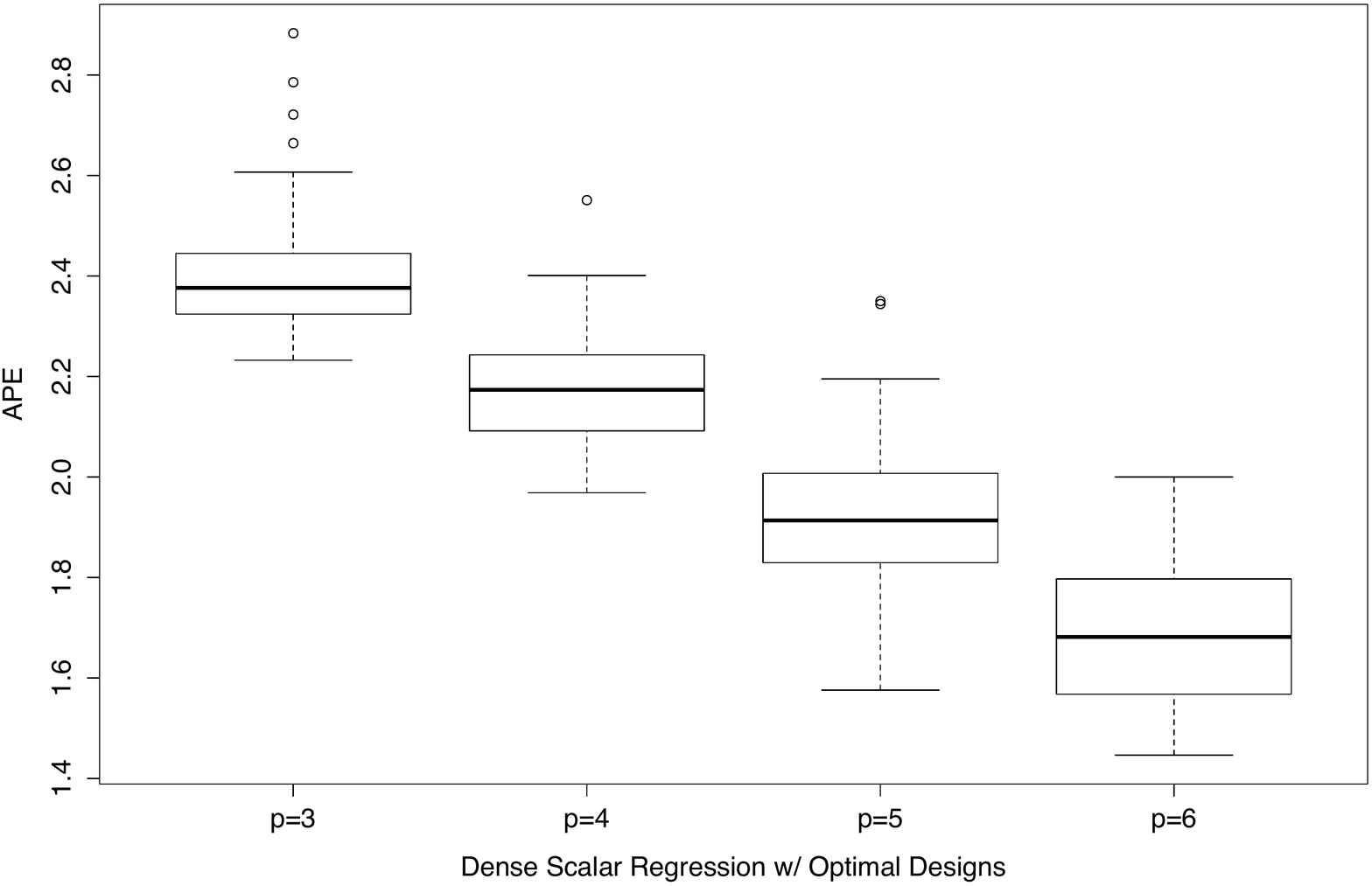}
\includegraphics[scale=0.2]{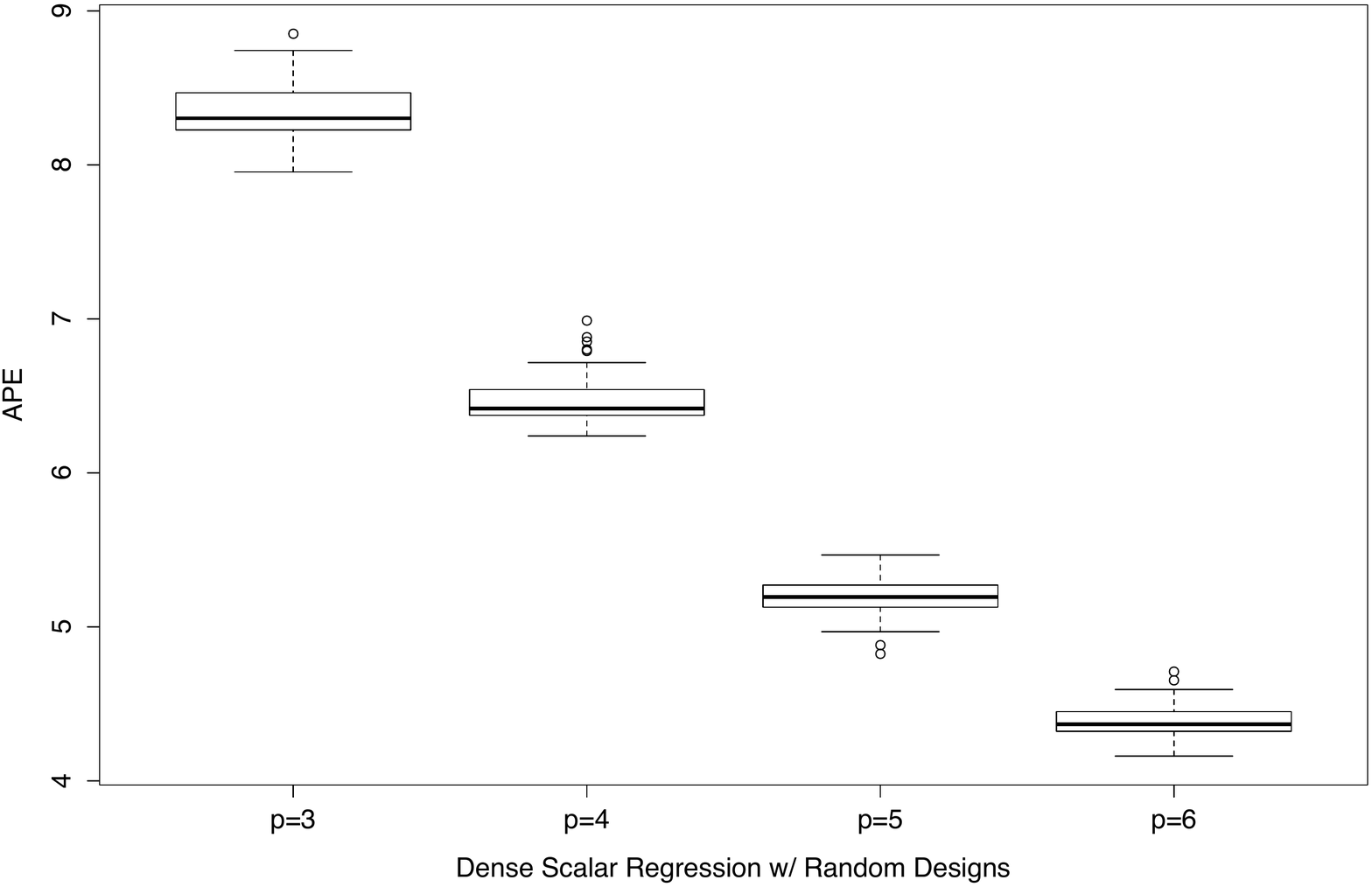}
\includegraphics[scale=0.2]{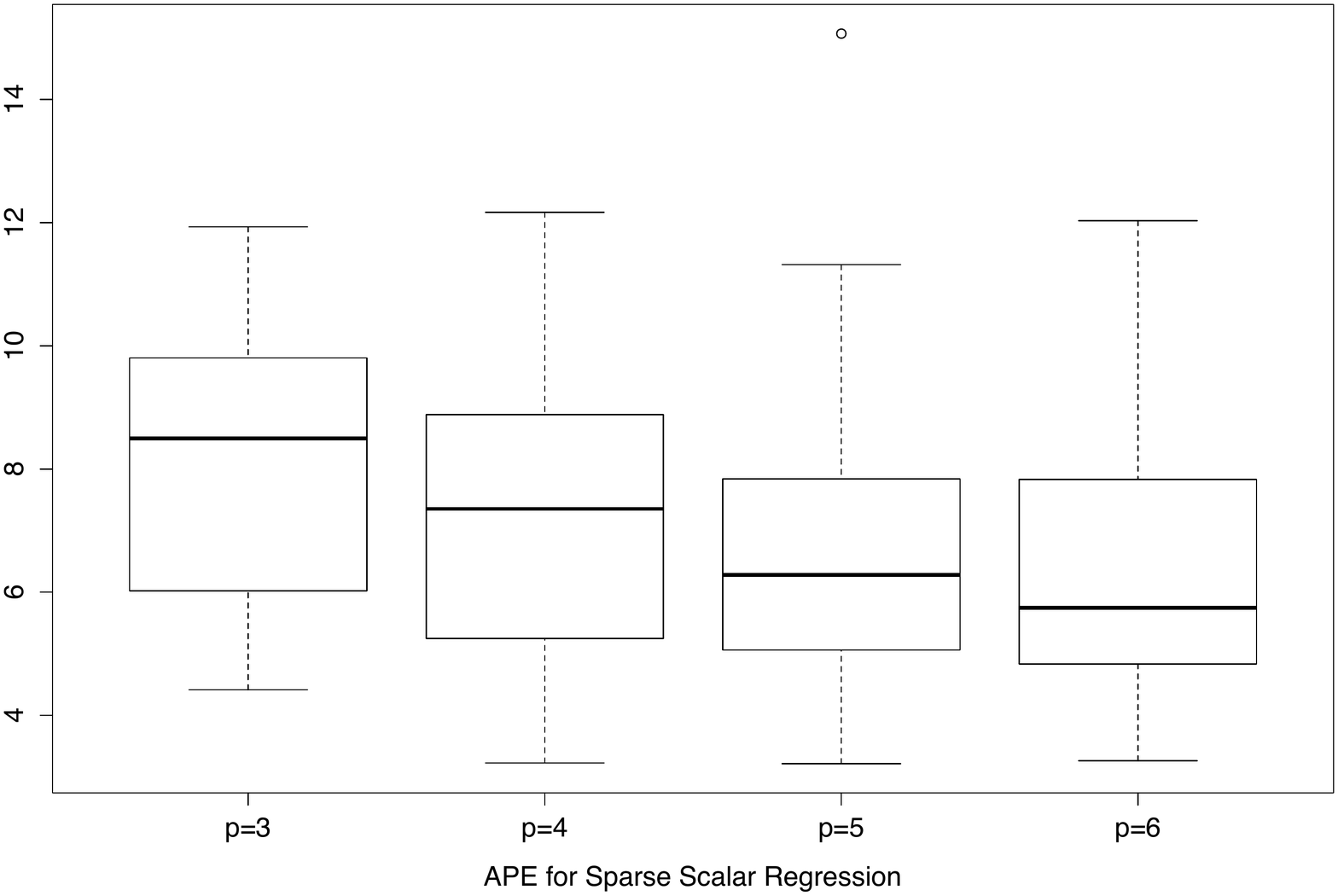}
\includegraphics[scale=0.2]{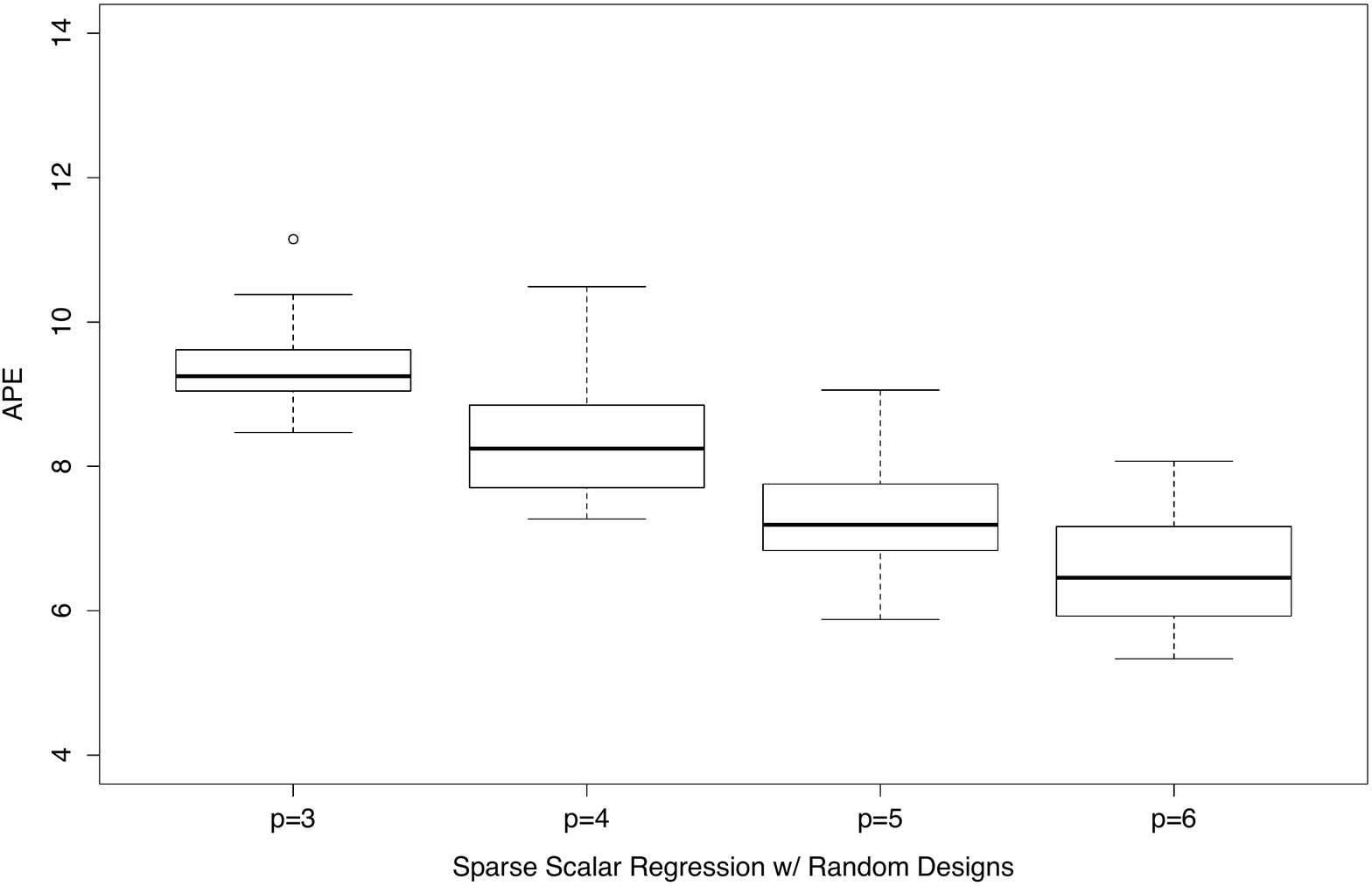}
\caption{Boxplots of APE from 100 simulation runs for scalar response regression for both dense and sparse scenarios with sequential search algorithm, with $p=3,4,5,6$. The first (second) row corresponds to the dense (sparse) scenario; the left (right) column has the results  for the optimal design (median performance random design).}
\end{figure}
\end{center}
\vspace{-1.5cm}

\sss{7.2 The Baltimore Longitudinal Study of Aging}
For the  Baltimore Longitudinal Study of Aging (BLSA)  we aim at identifying optimal designs for recovering Body Mass Index (BMI) profiles from sparse measurements and for  predicting a subject's systolic blood pressure (SBP) at old age.  To construct optimal designs, we use available pilot data that come from the longitudinal BLSA study, where measurements of BMI are sparse and irregularly spaced. To avoid bias due to censoring effects, we only include subjects with non-missing date of death and for whom age at death is above 70, and consider only the available  measurements that were taken within the age range from 45 to 70. We also exclude subjects who had less than four measurements. The response is taken to be the last SBP measurement before death. Within the study sample, 496 subjects met these criteria and were included in the analysis. A subset of the data, where measurements are connected by straight lines, is shown in Figure 1.

\begin{center}
\begin{figure}[H]\label{medflyspaghetti}
\centering
\includegraphics[scale=0.25]{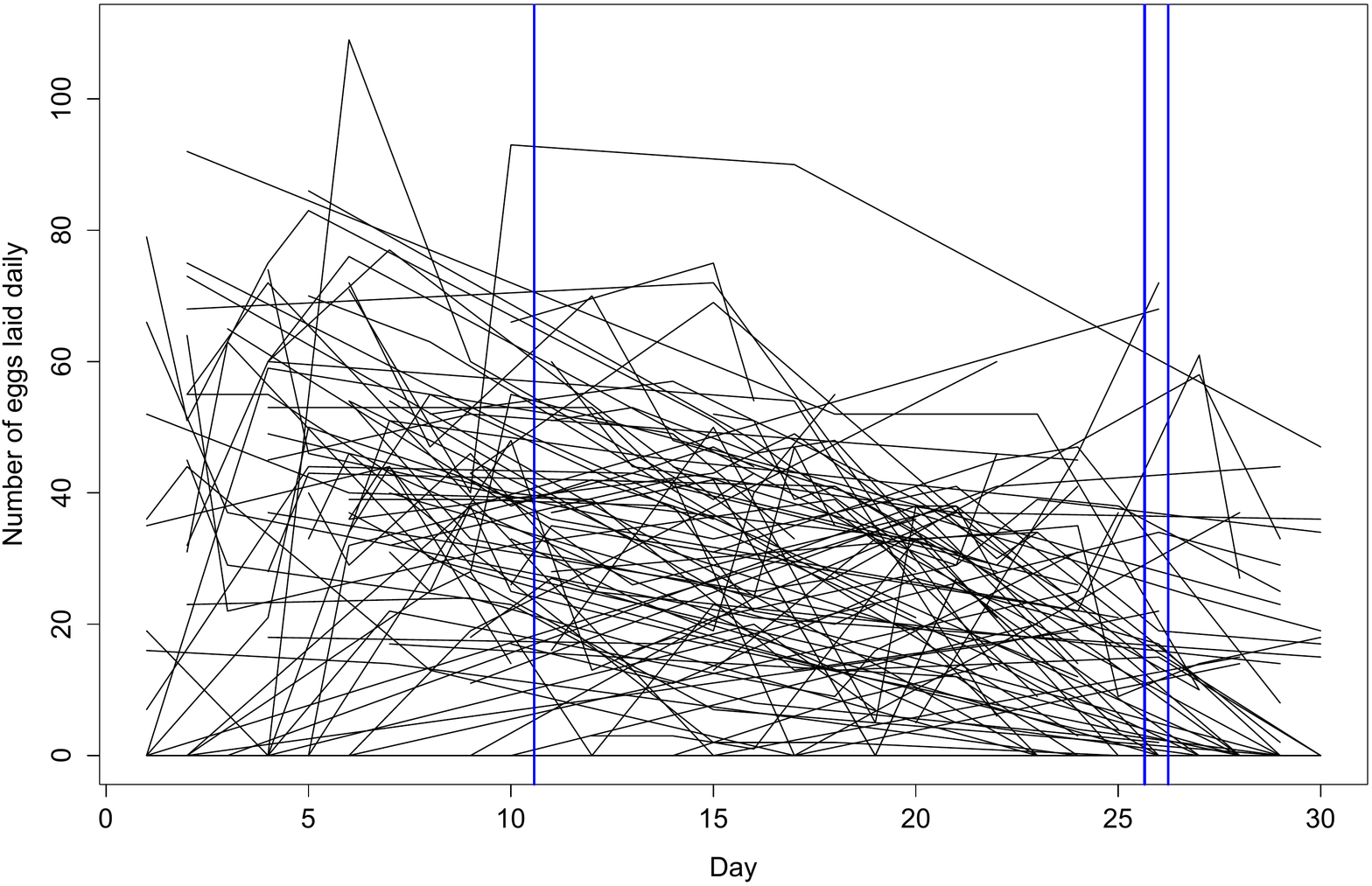}
\caption{Spaghetti plot for a subset of the training sample for predicting remaining egg-laying potential for female medflies from their egg-laying profiles. The vertical bars denote the locations of the optimal design points for $p=3$.}
\end{figure}
\end{center}
\vspace{-1cm}

A decisive difference between  these data and the previous data illustration is that the BMI trajectories that are assumed to generate the observed data are not available, as these pilot data are from a typical longitudinal study with inherently sparse and irregular measurement times.  Therefore, only indirect information is available for each subject about the underlying trajectory and only conditional inference about the trajectories is possible, conditioning on the available measurements for each subject \cp{yao:05:1}. This is the typical situation one faces when constructing optimal designs in the common situation where  the available pilot data are from a longitudinal study.  Nevertheless, the construction of optimal designs is still possible since they only depend on the covariance structure of the data, which can still be consistently estimated. To implement the proposed optimal designs for this situation, we use the modified cross validation method to select the ridge parameter.

Here the construction of optimal designs is intended for future data collection in longitudinal studies that will adopt the same  fixed optimal design for the subjects to be included in the subsequent study.
Due to the fact that the pilot data are coming  from a longitudinal study with random measurement locations, subjects included  in the pilot study will normally not have  measurements at the optimal design locations. As in addition their actual functional trajectories are unknown, it is not possible within the framework of the  BLSA study  to directly evaluate the performance of the optimal designs in terms of  recovering the BMI trajectories or predicting old age systolic blood pressure (SBP). Alternatively, we consider performance as measured by the coefficients of determination  in (\ref{integR2}) and (\ref{R2Y}). The selected optimal designs, along with their associated coefficients of determination,  are listed in Table 3 for $p=3,4,5$, for both prediction and trajectory recovery, where exhaustive search was used to determine these designs.

\vspace{-.5cm}
\begin{center}
\begin{figure}[H]\label{medflyp3}
\centering
\includegraphics[scale=0.25]{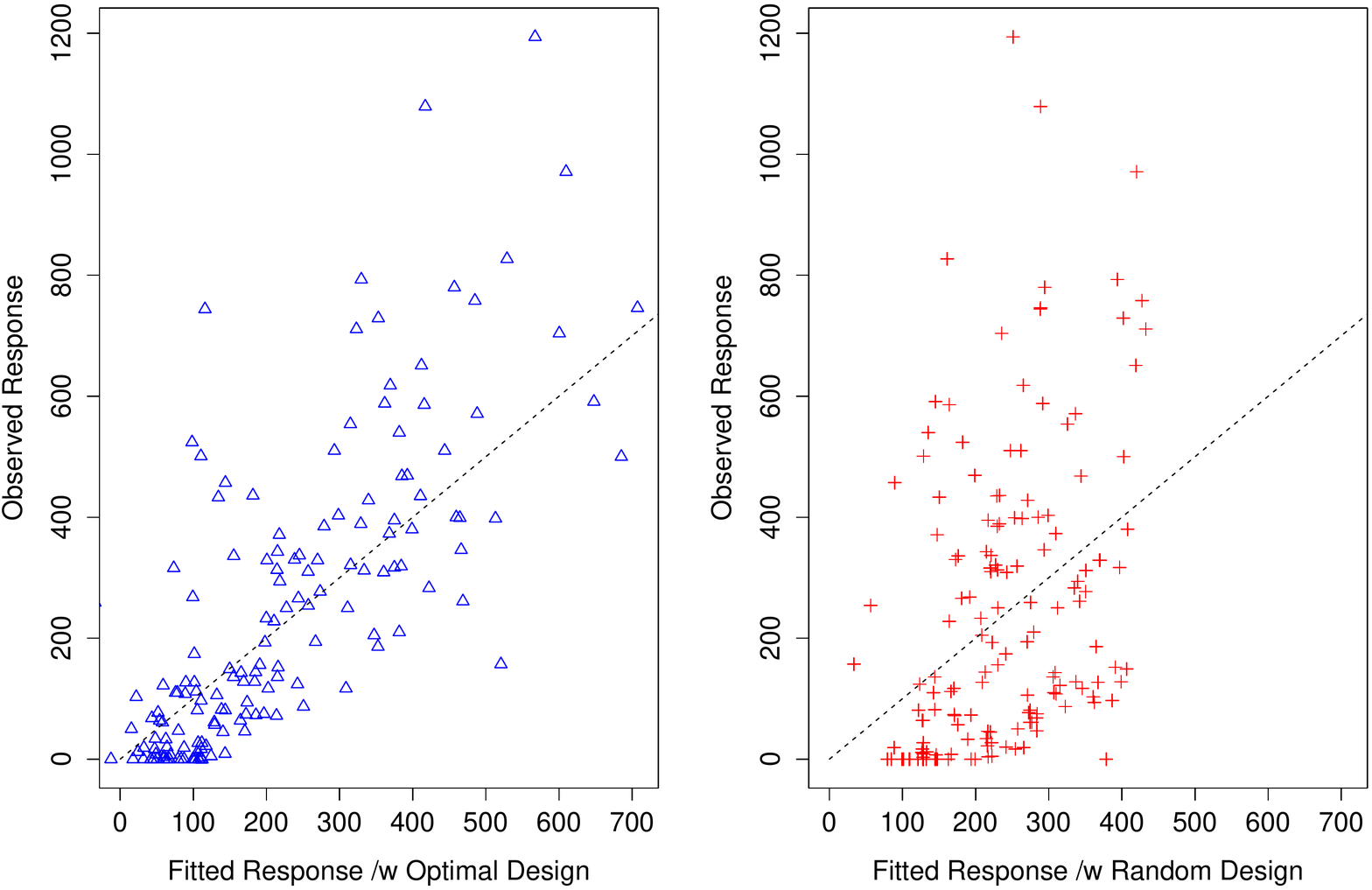}
\caption{Observed versus fitted (predicted) numbers of eggs that female medflies lay in their remaining lifetime when using $p=3$ design points in a functional linear regression model.  The predictions  for optimal designs are in the left panel and those for median performance random designs in the right panel. The black dotted  line is the 45 degree line in both graphs.}
\end{figure}
\end{center}

\vspace{-1cm}
The following findings are of interest for this and similar situations where one has longitudinal pilot data:
First, we  the designs for the prediction and the recovery task were found to differ  somewhat, especially in terms of older age measurements that are part of the optimal designs for trajectory recovery if $3  \le p \le  4$, but are somewhat less relevant for predicting the systolic blood pressure response at 70. Therefore, optimal designs for old age SBP prediction from BMI will not require measurements at older ages.
Second,  the constructed optimal designs for the same target quantity are consistent in the sense that as $p$ increases, additional design points are added while the previously selected design points are still viable so that selected optimal designs to some extent form a nested sequence, even when exhaustive search is used to determine these designs.   Third, the design points for trajectory recovery tend to be more evenly distributed than those for response prediction.
In other words,  design points that are more or less uniformly distributed across the age range seem to achieve the goal of trajectory recovery well.

The latter point was also seen in the bicycle sharing data, which are described in Online Supplement A.6. The reason behind this might  be that the random variation in trends or curvature across subjects is relatively uniform in these examples (see  Figure 1).
The optimal design points for prediction of old age SBP are generally clustering around earlier ages, which is potentially of interest for public health and prevention.  Finally, the coefficients of determination for both trajectory recovery and prediction are increasing with $p$, showing better performance with increasing number of design points, as was also seen in the simulations.  To select the optimal designs together with the optimal $p$, one can use a suitable penalty for larger  $p$ that might reflect  the data collection cost in specific applications. 

\begin{table}[H]
\centering{\caption{Optimal Designs  (Ages Where Measurements are Placed) and Associated Coefficients of Determination for Different Numbers of Design Points for the BLSA Data, for Recovering Body Mass Index (BMI) and for Predicting Systolic Blood Pressure (SBP).}} \label{BLSAresults}
\label{simsetting}
\vspace{.1cm}
\begin{center}
\begin{tabular}{|c|c|c|}
  \hline
 \   & Designs for BMI Trajectory Recovery ($R_{X}^2$) & Designs for SBP Prediction ($R_{Y}^2$) \\
 \hline
  $p=3$ & 46.5, 58, 68 (0.951) & 45, 45.5, 53 (0.863) \\
  $p=4$ & 46, 56.5, 62.5, 70 (0.973) & 45, 45.5, 52, 52.5 (0.874) \\
  $p=5$ & 47, 47.5, 60.5, 61.5, 70 (0.989) & 45, 45.5, 53, 53.5, 68.5 (0.886) \\
  \hline
\end{tabular}
\end{center}
\end{table}

Specifically for the BLSA study, comparing designs with different $p$, a general recommendation would be to  take the first measurement at around age 46 or 47, then the second within the age range 58 to 62, and finally a third at around age 70, if $p=3$ and BMI trajectory recovery is the objective. For predicting the systolic blood pressure, optimal designs are more concentrated in the first half of the age range, and it would be sufficient to take measurements at age 45 , 45.5 and around age 53.

\ss{8. Theory}
We establish asymptotic consistency and rates of convergence for the proposed optimal designs. Assumptions (A1)-(A7) and the proof of the following main result are in Online Supplement A.4.  The rates of convergence that we report here  are for the case that the pilot study is longitudinal and generates sparse functional/longitudinal data. In the following,  $h_\mu$, $h_S$ and $h_R$ are bandwidths for local linear smoothers for the mean function, cross-covariance function and auto-covariance function, as specified in (\ref{llmean}), (\ref{llcrosscov}) and (\ref{llautocov}) in Online Supplement A.2.

\begin{thm*}
 \label{thm}
 Assume that (A1)-(A7) hold, and that criteria $R^2_X$ and $R^2_Y$ are locally concave around the optimal design $\mathbf{t}_X^*$ for trajectory recovery (\ref{optdestraj}) and $\mathbf{t}_Y^*$ for scalar response regression (\ref{optdespred}) respectively, where $\mathbf{t}_X^*$, $\mathbf{t}_Y^* \in \mathcal{T}(\delta_0)$, defined at (\ref{Td}). For $\theta_n=h_R^2+\{\log n/ (nh_R^2)\}^{1/2}$, and given fixed $p$, the estimated optimal designs for trajectory recovery $\wh{\mathbf{t}}_X^*$ given by (\ref{optdestraj}) and for scalar response regression $\wh{\mathbf{t}}_Y^*$ given by (\ref{optdespred}) satisfy
$$||\wh{\mb{t}}_X^* - \mb{t}_X^*||_p = O(\theta_n^{1/2}) \ \ a.s. \ \ as\ \ n\rightarrow \infty$$
and
$$||\wh{\mb{t}}_Y^* - \mb{t}_Y^*||_p = O(\theta_n^{1/2}) \ \ a.s. \ \ as\ \ n\rightarrow \infty$$
where $||\mathbf{t}-\mathbf{s}||_p = \max_{1\leq j \leq p}|\mathbf{t}_{(j)}-\mathbf{s}_{(j)}|$, where $\mathbf{t}_{(j)}$, $\mathbf{s}_{(j)}$ are the $j$-th order statistics of designs $\mathbf{t}$ and $\mathbf{s}$. Here, we assume in addition that the smoothing bandwidths satisfy  $h_R^2 \lesssim h_\mu \lesssim h_R$, and $h_\mu \thicksim h_S$, where $a_n \lesssim b_n$ means $a_n=O(b_n)$, and $a_n \sim b_n$ means that $c\leq \frac{a_n}{b_n} \leq C$ for constants $0 < c < C < \infty$.
\end{thm*}

This result demonstrates  the consistency of the estimated optimal designs, including   rates of convergence to the true optimal designs. This provides theoretical justification for our methods. The rate of convergence $\theta_n$ is determined by the rate at which the bandwidth $h_R$ for smoothing the cross-covariance surface of the underlying smooth stochastic process converges to $0$.

The theorem is proved by  first establishing  uniform convergence of the optimization criteria (\ref{optdestraj}), (\ref{optpred}) with plugged-in estimators of the auxiliary quantities. The second step then is to prove the convergence of the estimated maximizer to the true maximizer. If the pilot study has dense and regular designs, one can apply cross-sectional covariance and mean estimation and by similar arguments provided here obtains analogous results as in the Theorem with the faster rate $\theta_n=n^{-1/2}$.

\ss{9. Discussion and Concluding Remarks}
We have developed a new method to obtain optimal designs for longitudinal data, by considering such data to be instances of functional data, i.e., by assuming that they are generated from an underlying (but unobservable) smooth stochastic process. This perspective makes it possible to take a pilot sample of sparsely observed longitudinal data and to construct optimal design points for future observations, where optimality refers to various criteria and loss functions. 

For the trajectory recovery task, the optimal designs are related to the shapes of the first few eigenfunctions that explain most of the variation. Specifically, the optimal design points are likely to cluster around areas where the variation in the underlying process across subjects is large, and that might correspond to areas 
where the first few eigenfunctions have peaks or valleys \cp{hall:06:3}, or generally areas where they are large in absolute value. On the other hand, for the scalar regression task, the optimal designs cannot be easily understood from the shape of the eigenfunctions, because these designs depend not only by the auto-covariance, but equally on the cross-covariance between responses and predictor trajectories.

Simulations and data analyses show that optimal designs can lead to substantial gains over random designs for both trajectory recovery and prediction. The constructed optimal designs thus provide guidance for the planning and collection of longitudinal data. The proposed method is conceptually straightforward and the estimating procedure is easy to implement. The method can be applied to a wide range of studies where dense measurements of underlying trajectories would be desirable but cannot be obtained due to various constraints. In various  engineering and medical applications, dense measurement designs are too expensive to obtain and therefore sparse designs need to be used. Some further extensions are discussed in online supplement A.8. 

While our focus in this paper is on optimal trajectory recovery and optimal prediction as two pertinent and important criteria, there are many other conceivable targets, such as predicting a functional response, or specific features of the underlying random trajectories, for example integrals or derivatives. It is also possible to select a mixed target criterion, targeting both trajectory recovery and optimal prediction.  One can then apply the same methodology as we describe here to find the optimal designs for such a mixed target criterion that blends various objectives, with relative weights given to each. For any given design, we can use the proposed criteria to evaluate its relative suitability for both prediction of a response as well as obtaining trajectory estimates in comparison with competing designs, for example by comparing the relative values of ARE (\ref{ARE}) and APE (\ref{APE}). This enables evaluation of the relative merits of potentially suboptimal designs that sometimes may be more convenient than optimal designs, or may be the result of extraneous constraints in data collection.

Our methods are also applicable in situations where in principle one can sample functional data densely but in future data collection only plans to sample at a few key locations. We find that the construction of optimal designs for longitudinal studies benefits in many ways from the adoption of a functional approach.

\vspace{-1cm}

\references

\newpage
\bc {\bf \Large Online Supplement} \ec

\bc {\it A.1 Best Linear Predictors} \ec
The best linear estimator for $X(t), t\in \mathcal{T}$ in (\ref{condexp1}) is derived as follows.\\
Given the observed longitudinal vector $\bf{U}$, assume that the best linear estimator for $X(t)$ takes the form
\begin{equation}
B(X(t)|\bf{U}) = a + {\bf{b}}^T {\bf{U}},
\end{equation}
where $a$ and $\bf{b}$ are constant scalar and vector respectively. In this way, $B(X(t)|\bf{U})$ minimizes the MSE, defined as
\begin{equation}\label{blpMSE}
\textit{MSE}(g) = \hbox{E} (X(t) - g({\bf{U}}))^2
\end{equation}
for any estimator $g(\cdot)$. Plugging in the linear presentation, we have 
$$\textit{MSE}(B(X(t)|{\bf{U}})) = \hbox{E} (X(t) - (a+{\bf{b}}^T \bf{U}) )^2.$$
Taking partial derivatives with respect to $a$ and $\bf{b}$ and setting to 0 yields the following two equations  for the constants,
\begin{equation}
\frac{\partial}{\partial a}{\textit{MSE}}(B(X(t)|{\bf{U}}) = 2 a + \hbox{E}(2{\bf{U}}^T{\bf{b}}) - 2\hbox{E}X(t) = 0,
\end{equation}
\begin{equation}
\frac{\partial}{\partial \bf{b}}{\textit{MSE}}(B(X(t)|{\bf{U}}) = 2 a \hbox{E}{\bf{U}} + 2\hbox{E}({\bf{UU}}^T{\bf{b}}) -2\hbox{E}(X(t) \bf{U}) = 0.
\end{equation}
Observing 
\be \nonumber \hbox{E}({\bf{U}\bf{U}}^T) &=& \hbox{Cov}({\bf{U}},{\bf{U}}) + (\hbox{E}{\bf{U}})(\hbox{E}{\bf{U}})^T=\hbox{Cov}({\bf{U}}) + \boldsymbol{\mu} \boldsymbol{\mu}^T,\\  \nonumber
\hbox{E}(X(t) {\bf{U}}) &=& \hbox{Cov}(X(t),{\bf{U}}) + \hbox{E}X(t) \hbox{E}{\bf{U}} = \hbox{Cov}(X(t),{\bf{U}}) + \mu(t) \boldsymbol{\mu},\ee
yields the solutions 
$$a = \mu(t) - \hbox{Cov}(X(t),{\bf{U}}) \hbox{Cov}({\bf{U}})^{-1}\boldsymbol{\mu}, \ \ \ {\bf{b}}=\hbox{Cov}({\bf{U}})^{-1} \hbox{Cov}({\bf{U}},X(t)).$$
Plugging in these expressions yields equation (\ref{condexp1}).

\bc {\it A.2 Local-linear Smoothing for Model Component Estimation} \ec
Let $K(\cdot)$ be a symmetric probability density function supported on $[-1,1]$ and $K_h(t)=(1/h)K(t/h)$, where $h$ is a  bandwidth or smoothing parameter. Then the mean function, auto-covariance function, cross-covariance surface and measurement error variance $\sigma^2$ are estimated as follows, applying the following smoothing steps.

\begin{enumerate}
\item The local-linear estimator of the mean function $\mu(t)$ is $\wh{\mu}_X(t)=\wh{a}_0$, where
\begin{equation} \label{llmean}
(\wh{a}_0,\wh{a}_1) = \hbox{argmin}_{a_0,a_1} \frac{1}{n} \sum_{i=1}^n \frac{1}{m_i} \sum_{j=1}^{m_i}\{U_{ij}-a_0-a_1(t_{ij}-t)\}^2 K_{h_\mu}(t_{ij}-t),
\end{equation}
with bandwidth $h_\mu$.

\item The  local-linear estimator of the cross-covariance function $C(t)$ is  $\wh{C}(t)=\wh{S}(t)-\wh{\mu}_X(t) \bar{Y}$, where $\bar{Y}=\frac{1}{n} \sum_{i=1}^n Y_i,\,$
$\wh{S}(t)=\wh{E}(X(t)Y)=\wh{a}_0$ and
\begin{equation} \label{llcrosscov}
(\wh{a}_0,\wh{a}_1) = \hbox{argmin}_{a_0,a_1} \frac{1}{n} \sum_{i=1}^n \frac{1}{m_i} \sum_{j=1}^{m_i}\{U_{ij}Y_i-a_0-a_1(t_{ij}-t)\}^2 K_{h_S}(t_{ij}-t),
\end{equation}
with bandwidth $h_S$.

\item The local-linear estimator of the auto-covariance function $\Gamma(s,t)$ is $\wh{\Gamma}(s,t)=\wh{R}(s,t) - \wh{\mu}_X(s)\wh{\mu}_X(t)$, where $\wh{R}(s,t)=\wh{E}(X(s)X(t))=\wh{a}_0$ and
\begin{equation} \label{llautocov}
(\wh{a}_0,\wh{a}_1,\wh{a}_2)=\hbox{argmin}_{a_0,a_1,a_2} \frac{1}{n} \sum_{i=1}^n [ \frac{1}{m_i(m_i-1)} \sum_{k \neq j} \{U_{ij}U_{ik}-a_0-a_1(t_{ij}-s)-a_2(t_{ik}-t)\}^2
\end{equation}
$$\times K_{h_R}(t_{ij}-s) K_{h_R}(t_{ik}-t)],$$
with bandwidth $h_R$.

\item Finally, to estimate $\sigma^2$, we first obtain an estimate of $V(t)=R(t,t)+\sigma^2$ by a local-linear estimator $\wh{V}(t)=\wh{a}_0$, where
\begin{equation} \label{lldiagcov}
(\wh{a}_0,\wh{a}_1) = \hbox{argmin}_{a_0,a_1} \frac{1}{n} \sum_{i=1}^n \frac{1}{m_i} \sum_{j=1}^{m_i}\{U_{ij}^2-a_0-a_1(t_{ij}-t)\}^2 K_{h_V}(t_{ij}-t).
\end{equation}
Then obtain  $\wh{\sigma}_X^2 = \frac{1}{|\mathcal{T}|} \int_\mathcal{T} \{\wh{V}(t)-\wh{R}(t,t) \} dt$. To account for boundary effects, one may tailor $\mathcal{T}$ by excluding the boundary area \cp{yao:05:1}.
\end{enumerate}

\bc {\it A.3 Maximizing Coefficients of Determination} \ec
In this section we show that for trajectory recovery and scalar response regression cases, minimizing either the MISE for trajectory recovery or the MSPE for regression is equivalent to maximizing the proposed coefficients of determination.\\

\no {\bf Trajectory Recovery.}
For trajectory recovery, by (\ref{condcondtraj}), the MISE as given in (\ref{MISE}) becomes
\begin{eqnarray}
\MISE(\mb{t}) &=& \int_\mathcal{T}E\{[X(t)-B(X(t)|\mb{U})]^2\}dt \nonumber \\
    &=& \int_\mathcal{T}E(X(t)-\mu(t)- \boldsymbol{\gamma}(t)^T\boldsymbol{\Gamma}_*^{-1}(\mb{U}-\boldsymbol{\mu}))^2dt \nonumber \\
    &=& \int_\mathcal{T}\{ \Var(X(t))+\boldsymbol{\gamma}(t)^T\boldsymbol{\Gamma}_*^{-1}\boldsymbol{\Gamma}_* \boldsymbol{\Gamma}_*^{-1}\boldsymbol{\gamma}(t)- 2E[(X(t)-\mu(t))(\mb{U}-\boldsymbol{\mu})^T\boldsymbol{\Gamma}_*^{-1}\boldsymbol{\gamma}(t)] \} dt \nonumber \\
    &=& \int_\mathcal{T}\{ \Var(X(t)) - \boldsymbol{\gamma}(t)^T\boldsymbol{\Gamma}_*^{-1}\boldsymbol{\gamma}(t)\} dt.
\end{eqnarray}
Since the overall variance of the underlying process does not involve the current design points $\mb{t}$, minimizing $\MISE(\mb{t})$ is equivalent to maximizing the term
\begin{equation}
\int_\mathcal{T} \boldsymbol{\gamma}(t)^T\boldsymbol{\Gamma}_*^{-1}\boldsymbol{\gamma}(t) dt.
\end{equation}
This is exactly the optimization criterion described in (\ref{optdestraj}). Note that the fact that $\int_\mathcal{T} \Var(X(t)) dt$ is constant implies the equivalence to maximizing $R_X^2$ in (\ref{integR2}).\\

\no {\bf Scalar Response Regression.}
Based on (\ref{betap}), the mean squared prediction error (MSPE) given by (\ref{prederror}) becomes
\begin{eqnarray}
E [ Y-B\{ E(Y|X)|\mb{U}\}]^2 &=& E [Y-\mu_Y-\boldsymbol{\Gamma}_*^{-1} \mb{C}^T(\mb{U}-\mb{\mu}) ]^2 \nonumber \\
    &=& \Var(Y) + \mb{C}^T\boldsymbol{\Gamma}_* \boldsymbol{\Gamma}_*^{-1}\boldsymbol{\Gamma}_*\mb{C}-2\mb{C}^T\boldsymbol{\Gamma}_*^{-1}\mb{C} \nonumber = \Var(Y) - \mb{C}^T\boldsymbol{\Gamma}_*^{-1}\mb{C}.
\end{eqnarray}
This implies the equivalence of minimizing MSPE to maximizing the criterion given in (\ref{optpred}).

\bc {\it A.4 Asymptotic Results: Assumptions and Proof of Theorem} \ec
Following \citet{li:10}, let $\eta_{nk}=\left( n^{-1}\sum_{i=1}^n m_i^{-k} \right)^{-1}$, and assume

(A1) For some constants $0<m<M<\infty$, $m \leq f_T(t) \leq M$ for all $t \in \mathcal{T}$. Further, $f_T$ is differentiable with a bounded derivative.

(A2) The kernel function $K(\cdot)$ is a symmetric probability density function on $[-1,1]$, and is of bounded variation on $[-1,1]$. Define $v_2 = \int_{-1}^1 t^2 K(t)dt$.

(A3) $\mu(\cdot)$ is twice differentiable and the second derivative is bounded on $\mathcal{T}$.

(A4) All second-order partial derivatives of $\Gamma(s,t)$ exist and are bounded on $\mathcal{T}^2$.

(A5) $E(|e_{ij}|^{\lambda_\mu})<\infty$ and $E(\sup_{t\in \mathcal{T}}|X(t)|^{\lambda_\mu})<\infty$ for some $\lambda_\mu \in (2,\infty)$; $h_\mu \rightarrow 0$ and $(h_\mu^2+h_\mu/\eta_{n1})^{-1}(\log n/n)^{1-2/\lambda_\mu} \rightarrow 0$ as $n \rightarrow \infty$.

(A6) $E(|e_{ij}|^{2\lambda_R}) < \infty$ and $E(\sup_{t\in\mathcal{T}}|X(t)|^{2\lambda_R})<\infty$ for some $\lambda_R \in (2,\infty)$; $h_R \rightarrow 0$ and $(h_R^4+h_R^3/\eta_{n1}+h_R^2/\eta_{n2})^{-1}(\log n/n)^{1-2/\lambda_R} \rightarrow 0$ as $n \rightarrow \infty$.

(A7) $E(|e_{ij}|^{2\lambda_V})<\infty$ and $E(\sup_{t \in \mathcal{T}}|X(t)|^{2\lambda_V})<\infty$ for some $\lambda_V \in (2,\infty)$; $h_V \rightarrow 0$ and $(h_V^2+h_V/\eta_{n1})^{-1}(\log n/n)^{1-2\lambda_V} \rightarrow 0$ as $n \rightarrow \infty$.

\vspace{0.2in}
\no {\it Proof of the Theorem.}
Based on the assumptions (A1)-(A7) and the smoothing methods we use (see (\ref{llmean}), (\ref{llautocov}), (\ref{llcrosscov})) for the estimates of $\mu(t)$, $\Gamma(s,t)$, $C(t)$, uniform convergence and rates of convergence are shown in \cp{li:10}. 
If $h_R^2 \lesssim h_\mu \lesssim h_R$, $h_V+(\log n/n)^{1/3} \lesssim h_R \lesssim h_V^2n/\log n$, we have, as $n \rightarrow \infty$,
$$\sup_{t\in \mathcal{T}} |\wh{\mu}_X(t)-\mu(t)|=O(h^2_\mu+\{\log n / (n h_\mu)\}^{1/2}) \ \ a.s.,$$
$$\sup_{t\in \mathcal{T}} |\wh{C}(t)-C(t)|=O(h^2_S+\{\log n / (n h_S)\}^{1/2}) \ \ a.s.,$$
$$\sup_{s,t\in \mathcal{T}} |\wh{\Gamma}(s,t)-\Gamma(s,t)|=O(h^2_R+\{\log n / (n h_R^2)\}^{1/2}) \ \ a.s.$$
$$\wh{\sigma}_X^2 - \sigma^2 = O(h^2_R+\{\log n / (n h_R)\}^{1/2}) \ \ a.s.$$
Set $\theta_n=h^2_R+\{\log n / (n h_R^2)\}^{1/2}$. It is easy to see that $h^2_R+\{\log n / (n h_R)\}^{1/2}=O(\theta_n)$. Since $p$ is fixed and finite, $\mathcal{T}$ is compact, and that $X$ is $\mathcal{L}^2$ process, we have that $\Gamma(s,t)$ is finite for all $s,t\in\mathcal{T}$. As one assumption is  $h_R^2 \lesssim h_\mu \lesssim h_R$, for some $\delta_0>0$
$$\sup_{t\in \mathcal{T},\mb{t}\in\mathcal{T}(\delta_0)} |\wh{\boldsymbol{\gamma}}(t)^T \wh{\boldsymbol{\Gamma}}_*^{-1} \wh{\boldsymbol{\gamma}}(t)-\boldsymbol{\gamma}(t)^T \boldsymbol{\Gamma}_*^{-1} \boldsymbol{\gamma}(t)| = O(\theta_n) \ \ a.s.$$
as $n\rightarrow \infty$. So for all $\mb{t} \in \mathcal{T}(\delta_0)$, writing $g(\mb{t})=\int_\mathcal{T} \boldsymbol{\gamma}(t)^T \boldsymbol{\Gamma}_*^{-1} \boldsymbol{\gamma}(t) dt$, and $\wh{g}(\mb{t})=\int_\mathcal{T} \wh{\boldsymbol{\gamma}}(t)^T \wh{\boldsymbol{\Gamma}}_*^{-1} \wh{\boldsymbol{\gamma}}(t) dt$, we readily have
$$|\wh{g}(\mb{t}) - g(\mb{t})| \leq \int_\mathcal{T} |\wh{\boldsymbol{\gamma}}(t)^T \wh{\boldsymbol{\Gamma}}_*^{-1} \wh{\boldsymbol{\gamma}}(t)-\boldsymbol{\gamma}(t)^T \boldsymbol{\Gamma}_*^{-1}  \boldsymbol{\gamma}(t)| dt =O(\theta_n) \ \ a.s.,$$
and therefore,
$$\sup_{\mb{t} \in \mathcal{T}(\delta_0)} |\wh{g}(\mb{t}) - g(\mb{t})| \xrightarrow{n \rightarrow \infty} 0 \ \ a.s.$$
For trajectory recovery, the true optimal design is $\mb{t}^*_X=\arg\max_{\mb{t}\in \mathcal{T}(\delta_0)} g(\mb{t})$, and its estimate is $\wh{\mb{t}}^*_X=\arg\max_{\mb{t}\in \mathcal{T}(\delta_0)} \wh{g}(\mb{t})$.
We claim that $|g(\mb{t}_X^*)-\wh{g}(\wh{\mb{t}}_X^*)| = O(\theta_n)$ a.s., as when

$g(\mb{t}_X^*) \geq \wh{g}(\wh{\mb{t}}_X^*)$,
$0 \leq g(\mb{t}_X^*)-\wh{g}(\wh{\mb{t}}_X^*) \leq g(\mb{t}_X^*) - \wh{g}(\mb{t}_X^*) = O(\theta_n) \ \ a.s.$, and

$g(\mb{t}_X^*) \leq \wh{g}(\wh{\mb{t}}_X^*)$,
$0 \leq \wh{g}(\wh{\mb{t}}_X^*)-g(\mb{t}_X^*) \leq \wh{g}(\wh{\mb{t}}_X^*) - g(\wh{\mb{t}}_X^*)=O(\theta_n) \ \ a.s.$\\
By smoothness of $\Gamma(s,t)$, we have that $g: \mathcal{T}(\delta_0) \rightarrow \mathbb{R}$ is a continuous function. Now,
$$|g(\mb{t}_X^*)-g(\wh{\mb{t}}_X^*)| \leq |g(\mb{t}_X^*)- \wh{g}(\wh{\mb{t}}_X^*)| + |\wh{g}(\wh{\mb{t}}_X^*) - g(\wh{\mb{t}}_X^*)| = O(\theta_n) \ \ a.s.$$
which implies that
$$\wh{\mb{t}}_X^* \xrightarrow{n\rightarrow \infty} \mb{t}_X^* \ \ a.s.$$
Moreover, with a mean value $\boldsymbol{\xi}$,
$$|g(\mb{t}_X^*)-g(\wh{\mb{t}}_X^*)|=\frac{1}{2} (\wh{\mb{t}}_X^* - \mb{t}_X^*)^T g^{(2)}(\boldsymbol{\xi})(\wh{\mb{t}}_X^* - \mb{t}_X^*).$$
Using the assumption that $g$ is locally concave around $\mb{t}_X^*$ and the convergence derived above, we have
$$||\wh{\mb{t}}_X^* - \mb{t}_X^*||_p=O(\theta_n^{1/2}) \ \ a.s.$$
where $||\mathbf{t}-\mathbf{s}||_p = \max_{1\leq j \leq p}|\mathbf{t}_{(j)}-\mathbf{s}_{(j)}|$, where $\mathbf{t}_{(j)}$, $\mathbf{s}_{(j)}$ are the $j$-th order statistics of designs $\mathbf{t}$ and $\mathbf{s}$.

\noindent The proof for optimal regression designs is analogous.

\bc {\it A.5 Comparison between Optimal Designs and Random Designs} \ec
We compare selected optimal designs with ``random designs" where measurement locations are uniformly distributed over the domain, because in the sparse case there are no special designs that are representative for the data and equidistant designs are not commonly (in practice almost never) observed. For trajectory recovery of longitudinal data, it can be expected that the optimal designs are somewhat evenly spaced if the data show even distribution of variation across subjects, for example if  random curvature or trends are modest. 

This is the case for the BLSA data and the bike sharing data.  For scalar response regression, the situation is more complex, as the covariance function matters, and again there are no a priori special designs in the sparse random design case. Therefore it is best to base comparisons on the entire spectrum of designs that may be encountered, which is well approximated with the  space of random designs. Therefore the best comparison is indeed with ``random designs" to get a general idea of how the selected optimal designs work.

\bc {\it A.6 Additional Example: Bike Sharing} \ec
Bike sharing systems increasingly replace traditional bike rentals. In such systems the entire process of rental and return is automatized and data about the rentals are automatically generated.  The data we use are described in \ci{Fanaee:13} and include records of the hourly number of bike rentals for 2011 and 2012 from the Capital Bikeshare system, Washington D.C. We consider the trajectories corresponding to hourly numbers of bike rentals for all Saturdays and aim at finding the optimal design points to recover these bike rental trajectories. 

The bike rental data are densely and regularly measured, with 24 measurements per day, of which we randomly chose $3$ to $5$ measurements to create sparse designs, thus rendering the pilot data sparse.  The ridge parameter needed for implementation of the method is obtained by modified cross-validation. We focus on Saturday bike rentals and compare the recovered trajectories based on the selected optimal designs that are constructed from the sparsified pilot data with the actually observed curves. To assess  the performance of the proposed optimal designs in comparison to random designs, we include data for $80$ days in the training and for $22$ days in the testing sample. Figure 6 shows the Spaghetti plot for the training sample. The goal is to choose optimal design points for designs with $p=3$ measurements.
\begin{center}
\begin{figure}[H]\label{bikesharespaghetti}
\centering
\includegraphics[scale=0.35]{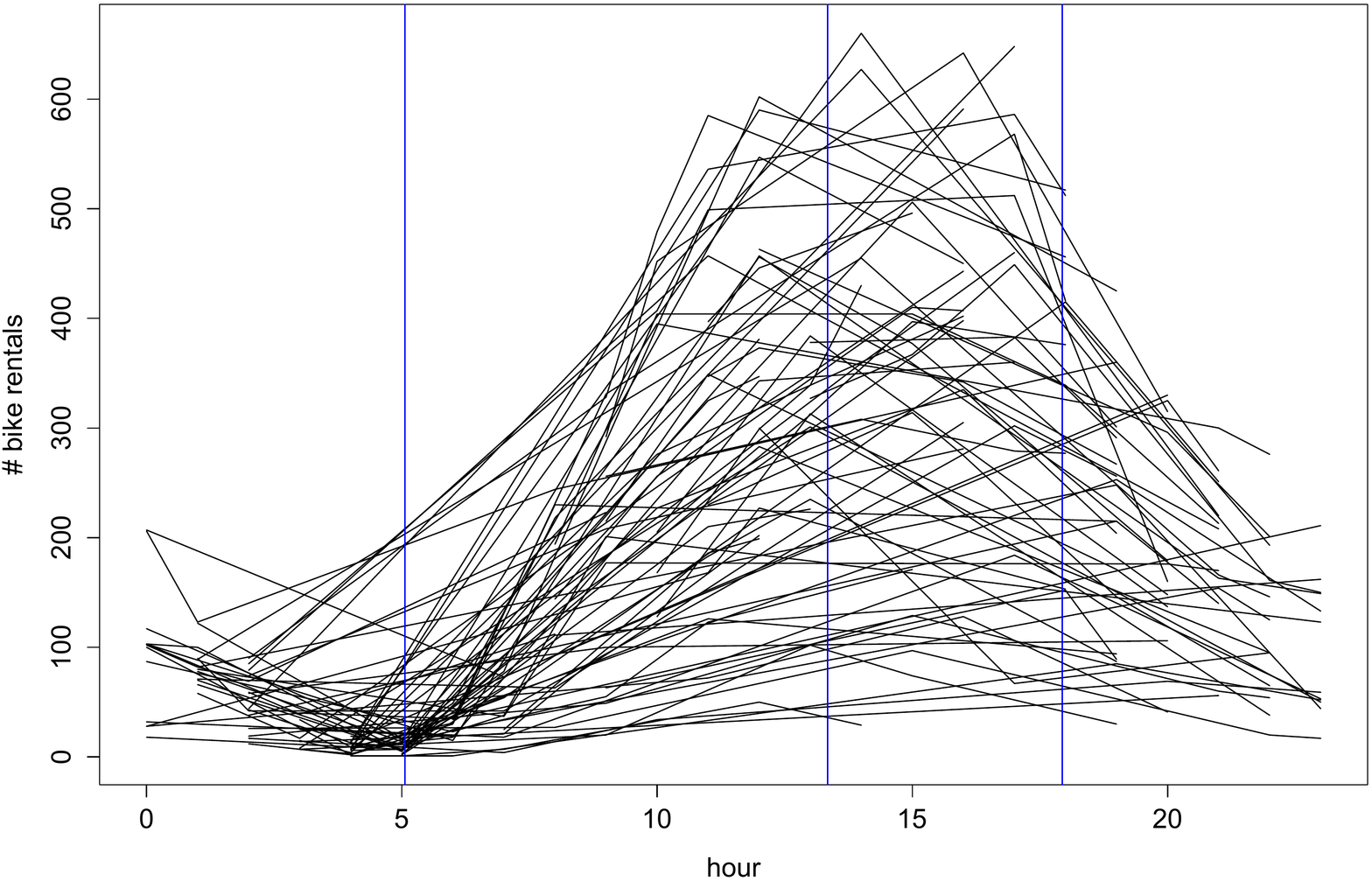}
\caption{Spaghetti plot for training sample of bike sharing data. The blue vertical bars denote the locations of the optimal design points for $p=3$ for recovering daily bike rental trajectories.}
\end{figure}
\end{center}
\vspace{-1.5cm}

\begin{center}
\begin{figure}[H]\label{bikep3}
\centering
\includegraphics[scale=0.35]{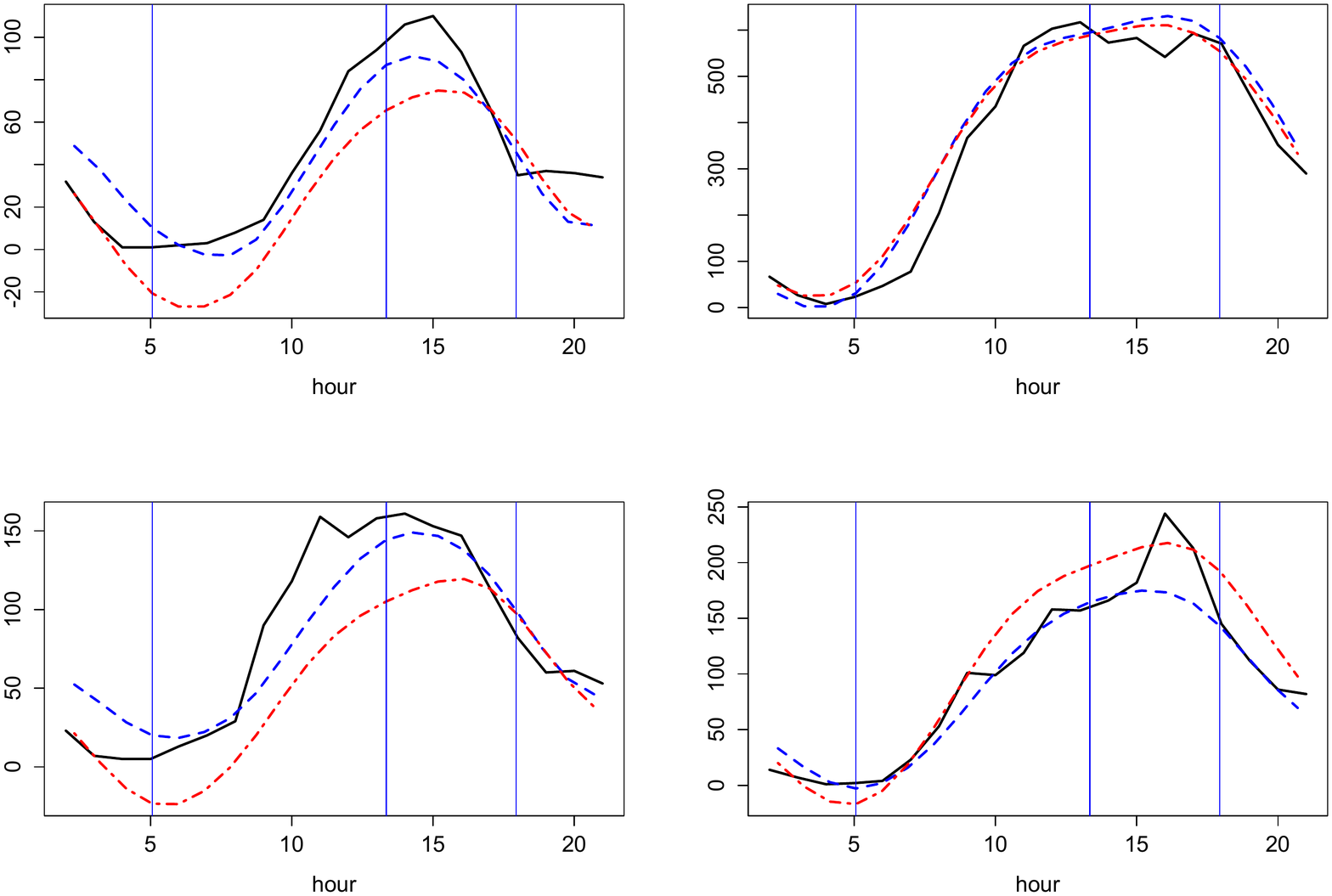}
\caption{Trajectory recovery results for four randomly chosen days from the test sample for the bike sharing data with sparse pilot measurements for $p=3$. Black curves stand for the underlying dense data. Blue dashed and red dotted curves are trajectory estimates using optimal designs and random designs, respectively. The  vertical blue lines are the locations for the optimal design points.}
\end{figure}
\end{center}
\vspace{-1.4cm}

Randomly selected subjects from the test set and their recovered trajectories are illustrated in Figure 7.
 Relative ARE (\ref{ARE}) for  the test sample using optimal designs is $0.264$, compared to $0.321$ for random designs of median performance for $p=3$.
 In Figure 7, the optimal designs are also seen to perform better than median performance random designs.
 The selected optimal design points are located at 5am, 1:30pm and 6pm during the day. These three points are located around time points corresponding to morning valley, noon peak and afternoon decrease of the bike renting profile for each day. These three points thus likely reflect key features of the daily bike renting profiles and therefore it is not surprising that their locations constitute the optimal design.

\bc {\it A.7 Additional Discussion on Simulation Studies} \ec
In our simulation studies, we also compared sequential design selection and exhaustive search and the effect of the ridge parameter $\sigma^2_{\rm{new}}$ on the performance of optimal designs. The results are summarized in Table 3 and Figure 8.
\begin{table}[H]
\caption{Investigating Sequential Optimization on Optimal Designs in Simulations, in Terms of Mean ARE  and Relative ARE (\ref{ARE}) (in brackets) for Trajectory Recovery and APE  and Relative APE (\ref{APE}) (in brackets) for Response Prediction in a Functional Linear Model based on 100 simulations.}
\begin{center}
\label{simtableSeq}
\begin{tabular}{|c|cc|cc|c}
  \hline
    & \multicolumn{2}{c}{Mean ARE (ARE${}^*$)} \vline & \multicolumn{2}{c}{Mean APE (APE${}^*$)} \vline \\
  \hline
   Sequential &   Dense   &   Sparse  & Dense & Sparse\\
   \hline
  $p=2$ & $1.74(.092)$ & $1.83(.101)$ & $2.91(.234)$ & $9.23(.745)$\\
   \hline
  $p=3$ & $1.37(.072)$ & $1.59(.088)$ & $2.4(.193)$ & $7.85(.630)$\\
   \hline
  $p=4$ & $1.02(.054)$ & $1.34(.074)$ & $2.18(.175)$ & $7.48(.601)$\\
  \hline 
  $p=5$ & $1.01(.053)$ & $1.32(.073)$ & $1.91(.154)$ & $6.71(.542)$\\
  \hline
  $p=6$ & $0.91(.048)$ & $1.24(.069)$ & $1.68(.136)$ & $6.29(.508)$\\  
  \hline
\end{tabular}
\end{center}
\end{table}
\vspace{-0.5cm}

\begin{center}
\begin{figure}[H]\label{ridgeeffect}
\centering
\includegraphics[scale=0.25]{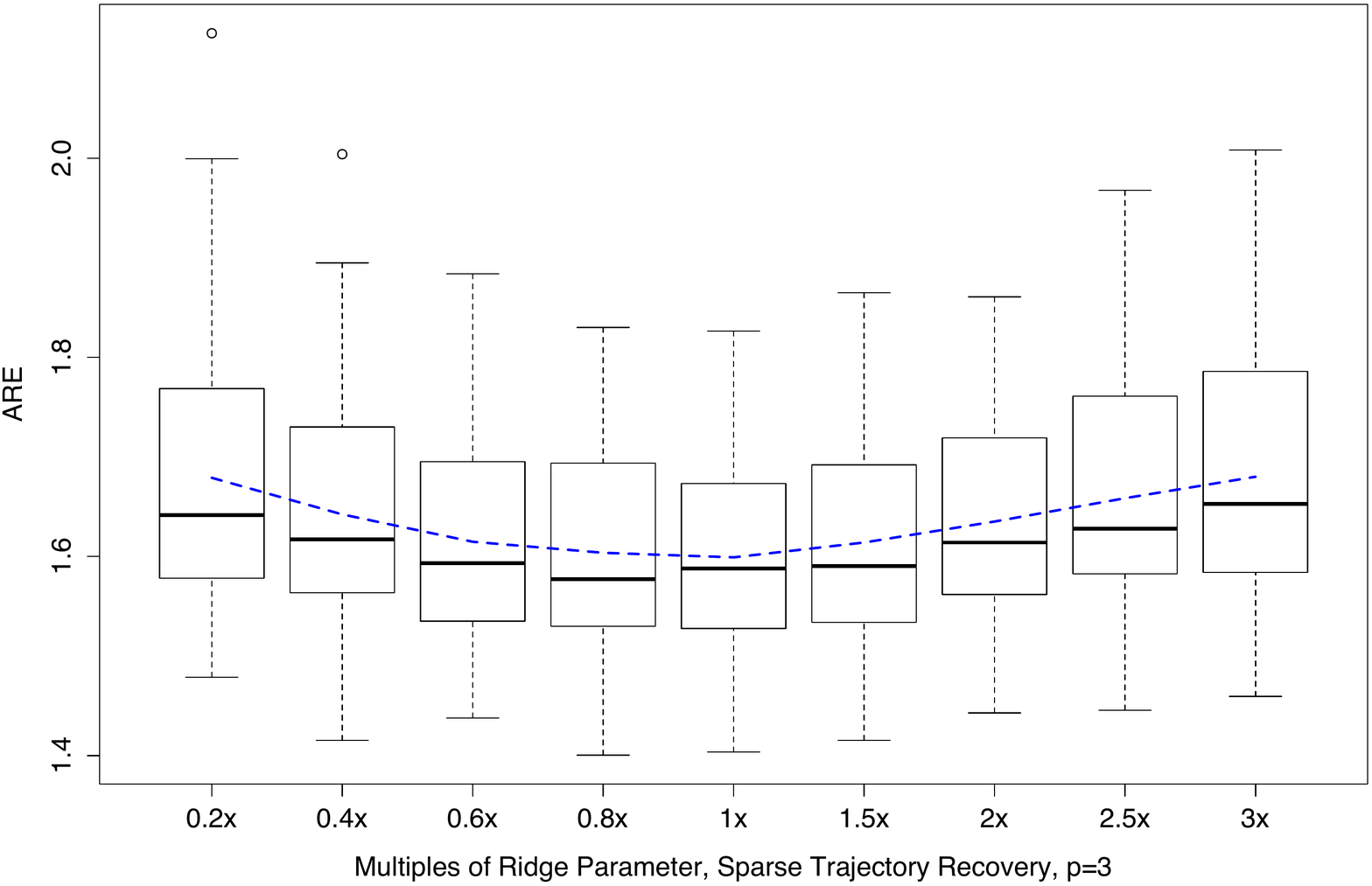}
\includegraphics[scale=0.25]{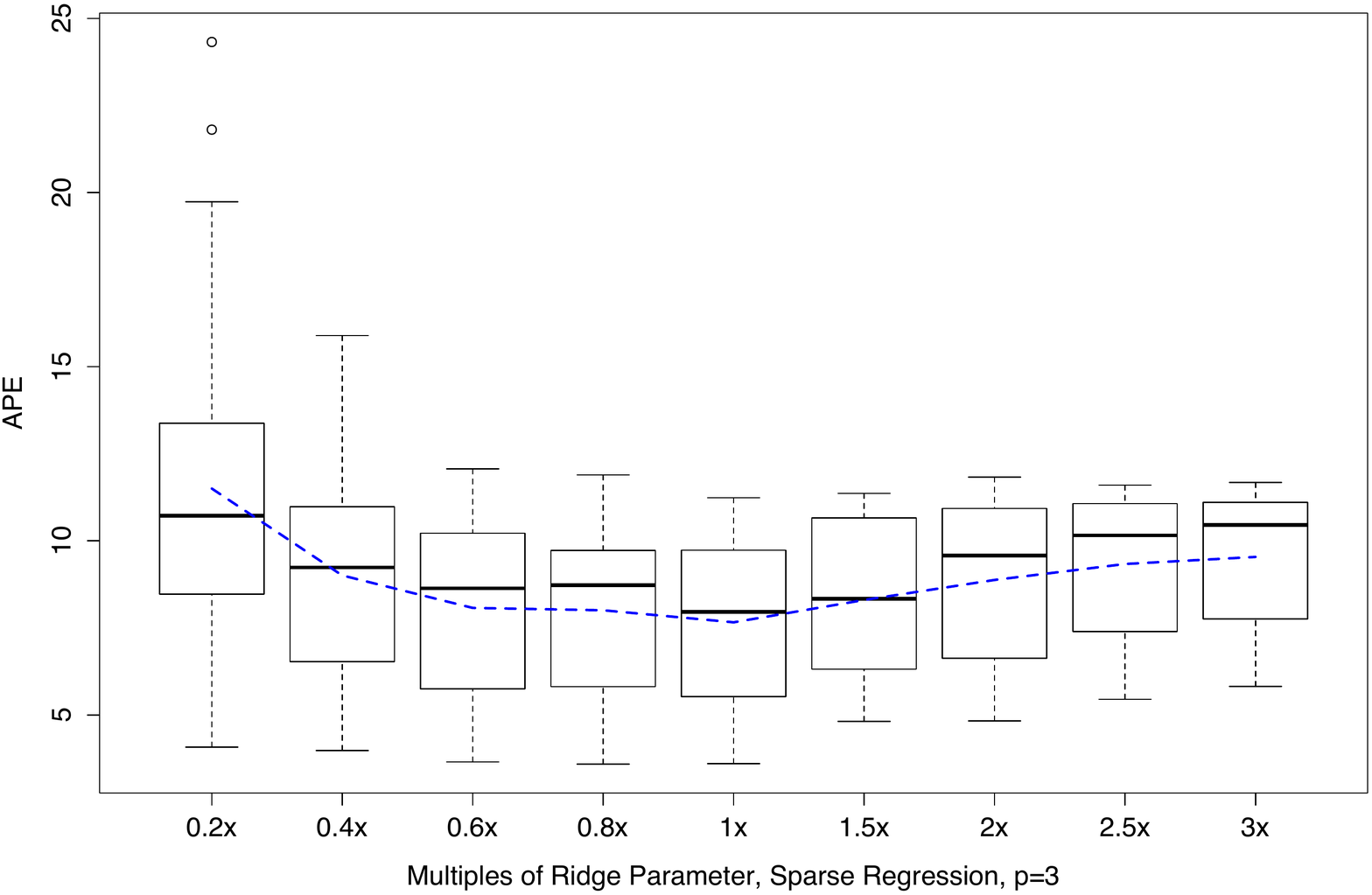}
\caption{Simulation results for the effect of the ridge parameter $\sigma^2_{\rm{new}}$ for 100 simulation runs. The left (right) panel depicts boxplots of ARE (APE) from sparsely observed simulation data by using multiples of the optimal ridge parameter selected by cross validation for trajectory recovery (scalar response regression) for $p=3$. Blue dotted lines denote the mean error across the different ridge parameters.}
\end{figure}
\end{center}
\vspace{-1.1cm}

Table 3 summarizes the simulation results of optimal designs from sequential selection algorithm for both dense and sparse scenarios with $p=2,3,4,5,6$ based on 100 simulations. The performance of sequential optimization is comparable to that of exhaustive search in all simulation scenarios, indicating that in the simulation scenario only a  minor loss in optimality occurs in exchange for a substantial reduction in computing time.

Ridge parameter also affects the performance of optimal designs.  Figure 7 shows ARE (APE) based on different multiples of selected ridge parameter via cross validation. For both simulation scenarios, the error is not very sensitive to the choice of ridge parameter, and similar results are observed for different $p$ as well as for densely observed data. In practice, the candidate set from which ridge parameters are selected can be chosen heuristically, where the selected outcome should not be located  at the boundary of the candidate set. The selection is relatively fast when  combined with the sequential search algorithm for optimal designs.

\bc {\it A.8 Additional Discussion} \ec

For some applications where sequential design selection is desirable,  given that our proposed method relies on estimates of mean and covariance  of the functional data, we could continuously update these estimates when additional measurements for one or more subjects become available.  This would be done by combining the new measurements with the current pilot data and then recalculating the optimal design. The resulting optimal designs would then be applied for the next batch of subjects, and one could iterate this scheme. Though numerically possible, we do not recommend updating the optimal design given a subset of measurements for a particular subject, because it is less practical to reschedule subject visits in longitudinal studies as would be advisable when the design has changed, and also mean and covariance surface are unlikely to change much when adding the measurements of a single subject or a subset of these measurements.

Another extension of our proposed method is to select optimal designs reaching a specified predictive accuracy threshold at a location in  the domain 
which is as early as possible, so that predictions of scalar outcomes or the remaining random trajectory can then be computed relatively early. This  would allow a longitudinal study to run for a shorter time, which can lead to faster results and cost savings. 

 The predictive accuracy threshold in this setting can be chosen as the coefficient of determination $R^2$ (subscripts $X$ and $Y$ suppressed for convenience), which is scale-invariant and applicable for both functional and longitudinal pilot studies. Suppose that the functional support is $\mathcal{T} = [0,T]$, and define $R^2(t)$ for $t\in (0,T]$ as the coefficient of determination attained by the optimal design, selected with the proposed method, but using data only on the time domain $[0,t]$. The function $R^2(t)$ increases with $t$ because the candidate set of designs becomes larger as $t$ increases. With a user-specified threshold parameter $\alpha \in (0,1)$, the final outcome is the optimal design that corresponds to the optimal design obtained from data on the interval $[0,t_\alpha]$, where
$$t_\alpha = \min\{t\in (0,T]: R^2(t) \geq (1-\alpha) R^2(T)\}.$$
This version of constructing optimal designs can be appealing for certain applications, since it could potentially save a lot of resources, with an acceptable decrease in predictive performance.

\bc {\it A.9 Additional Figures} \ec
\begin{center}
\begin{figure}[H]\label{simspaghetti}
\centering
\includegraphics[scale=0.25]{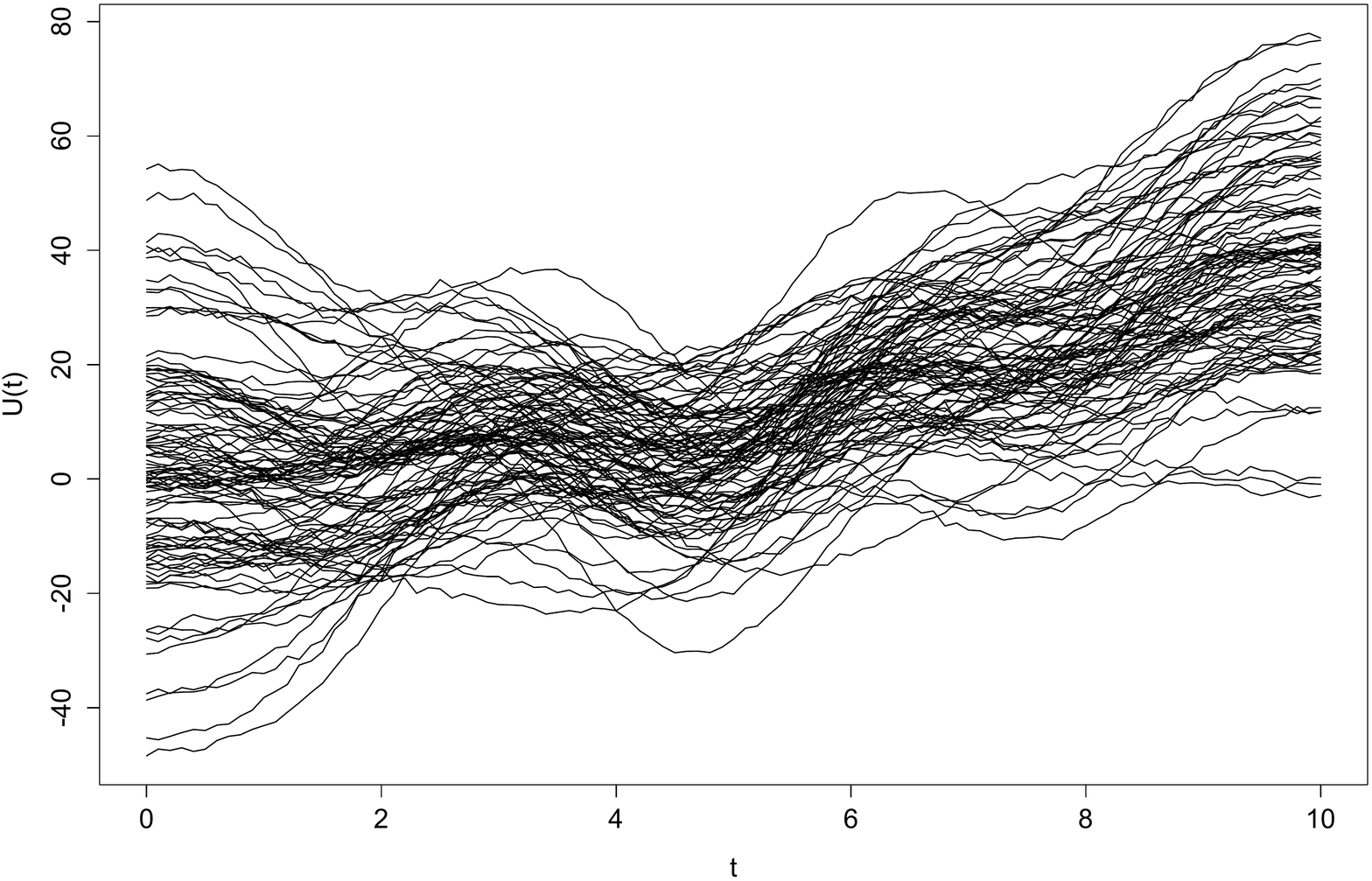}
\includegraphics[scale=0.25]{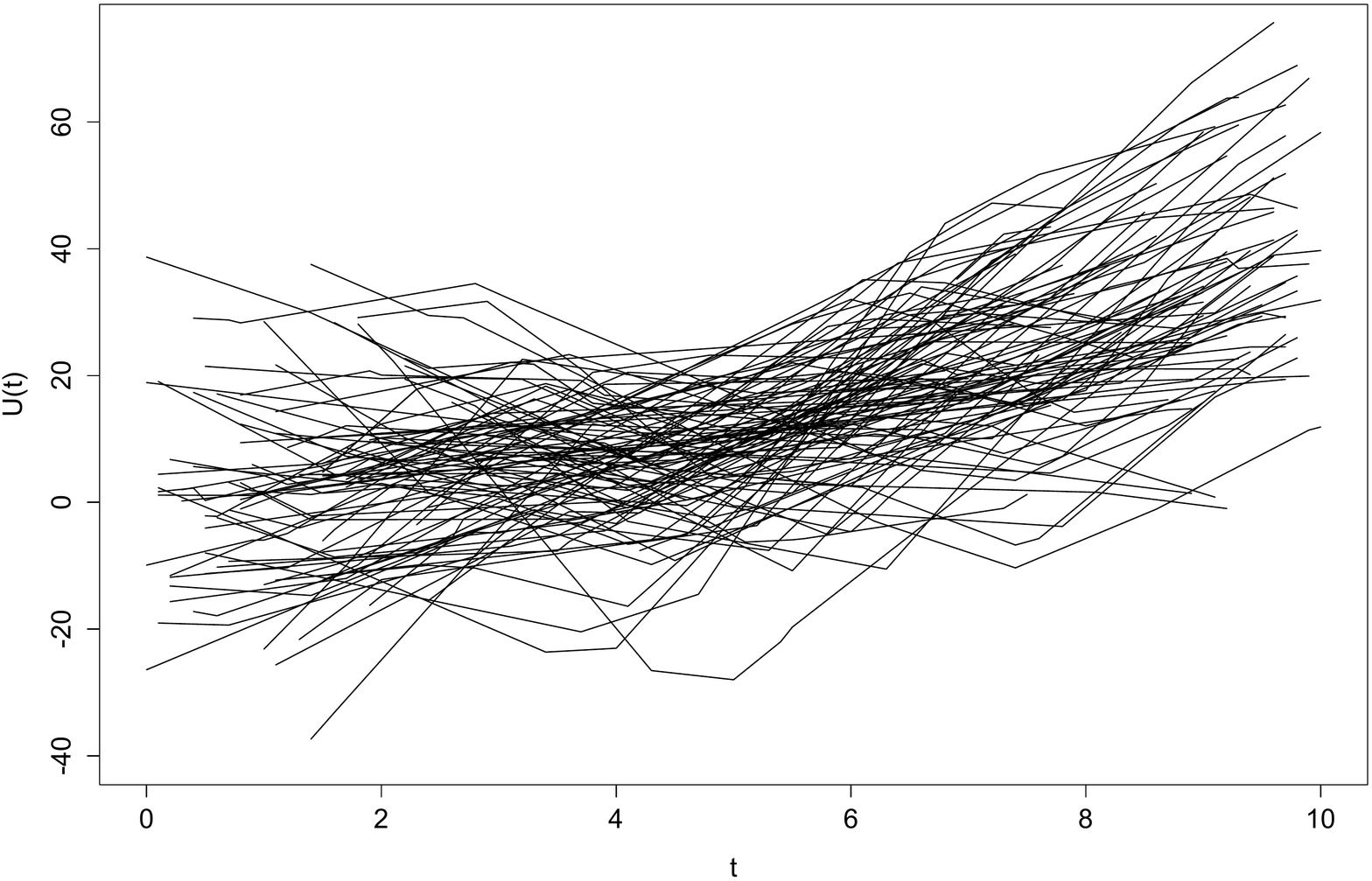}
\caption{Training samples for one simulation run. Left panel depicts a sample of size $n=100$ densely sampled  functional data (scenario 1)  and right panel is the spaghetti plot of a sample of 100 simulated training data based on a sparse functional sampling design (scenario 2).}
\end{figure}
\end{center}

\begin{center}
\begin{figure}\label{simTR3sparse}
\centering
\includegraphics[scale=0.32]{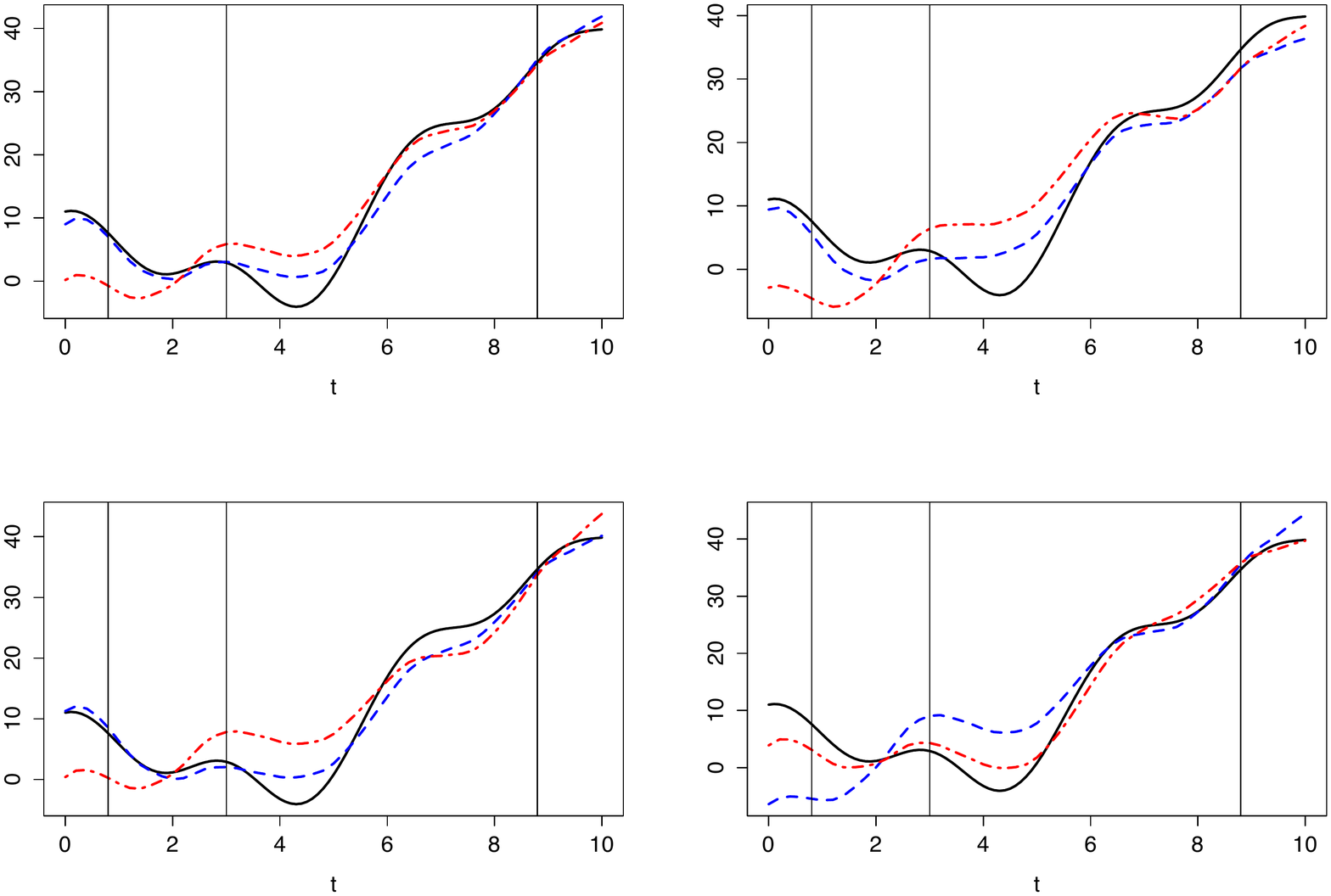}
\includegraphics[scale=0.32]{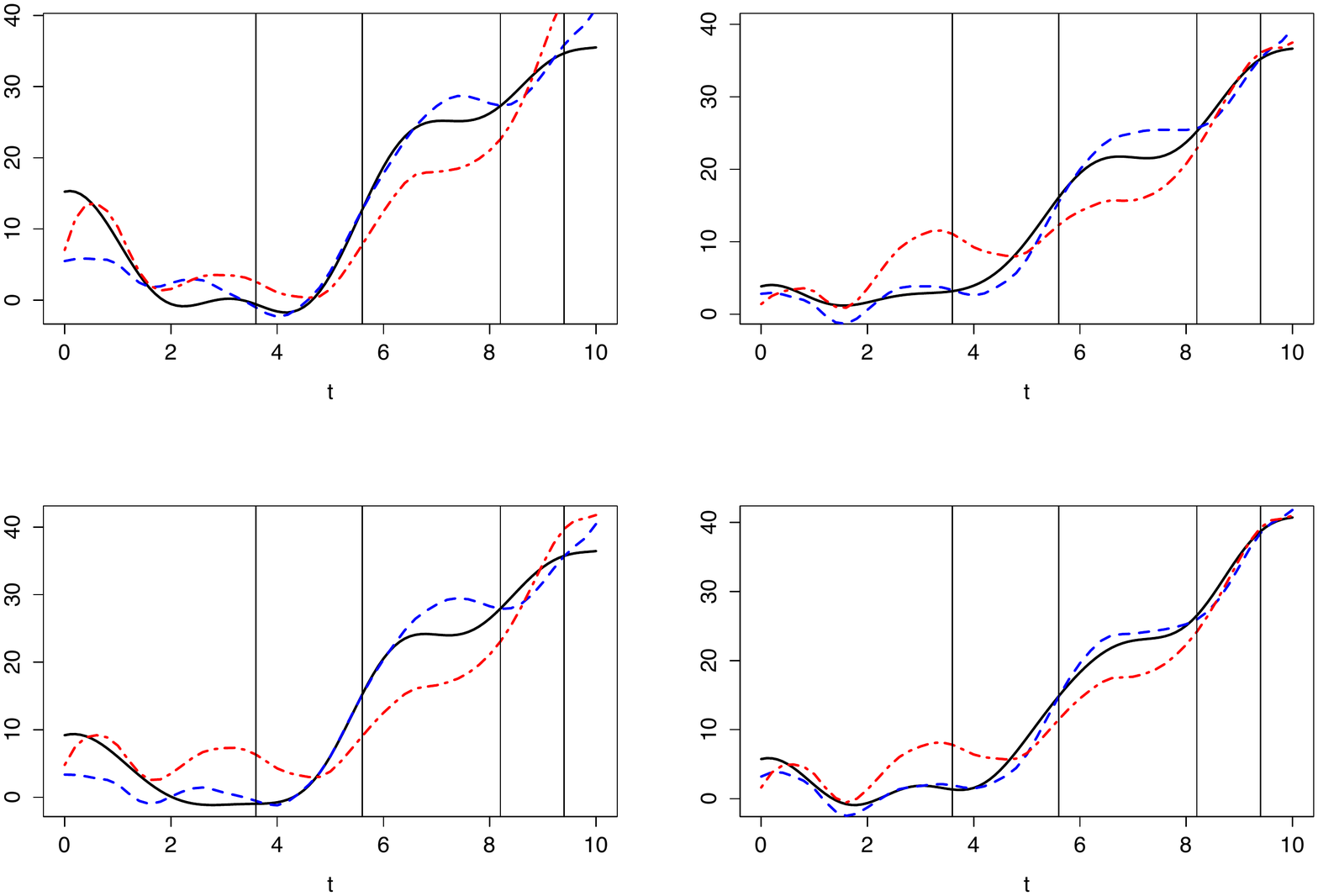}
\caption{Simulation results for sparsely sampled pilot data: Recovered trajectories for eight randomly chosen subjects from the test sample, based on 3 (upper four
panels)   or 4 (lower four panels) design points. Included are in each panel the true underlying curves (black solid curves), estimated trajectories from random designs with 3 (4) support points (red dash-dotted curves), for the random design with median performance, the estimated trajectories  for the optimal designs (blue dashed curves), and the optimal design point locations (black vertical lines).}
\end{figure}
\end{center}
\vspace{-1cm}
\begin{center}
\begin{figure}\label{simbox}
\centering
\includegraphics[scale=0.25]{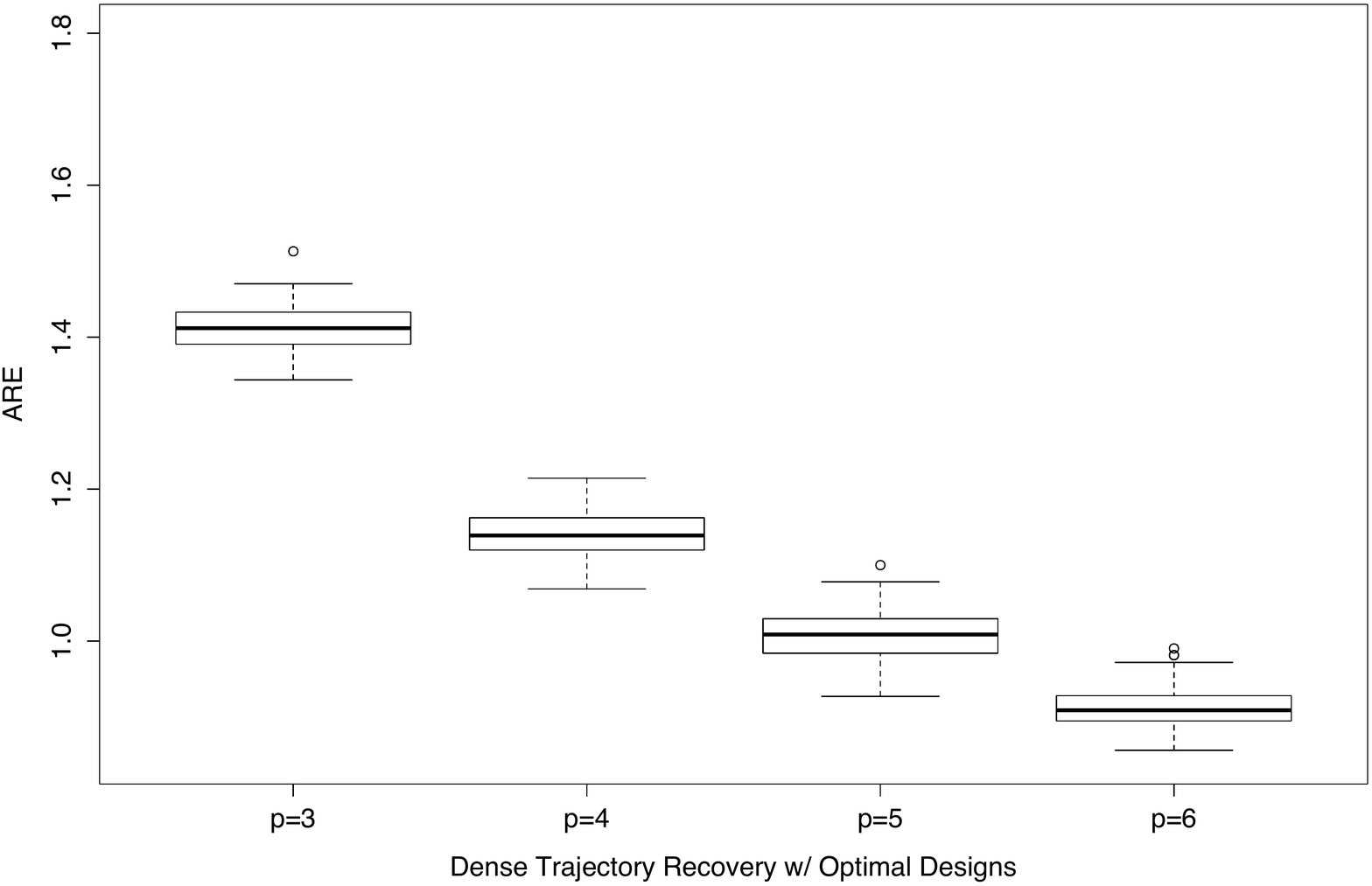}
\includegraphics[scale=0.25]{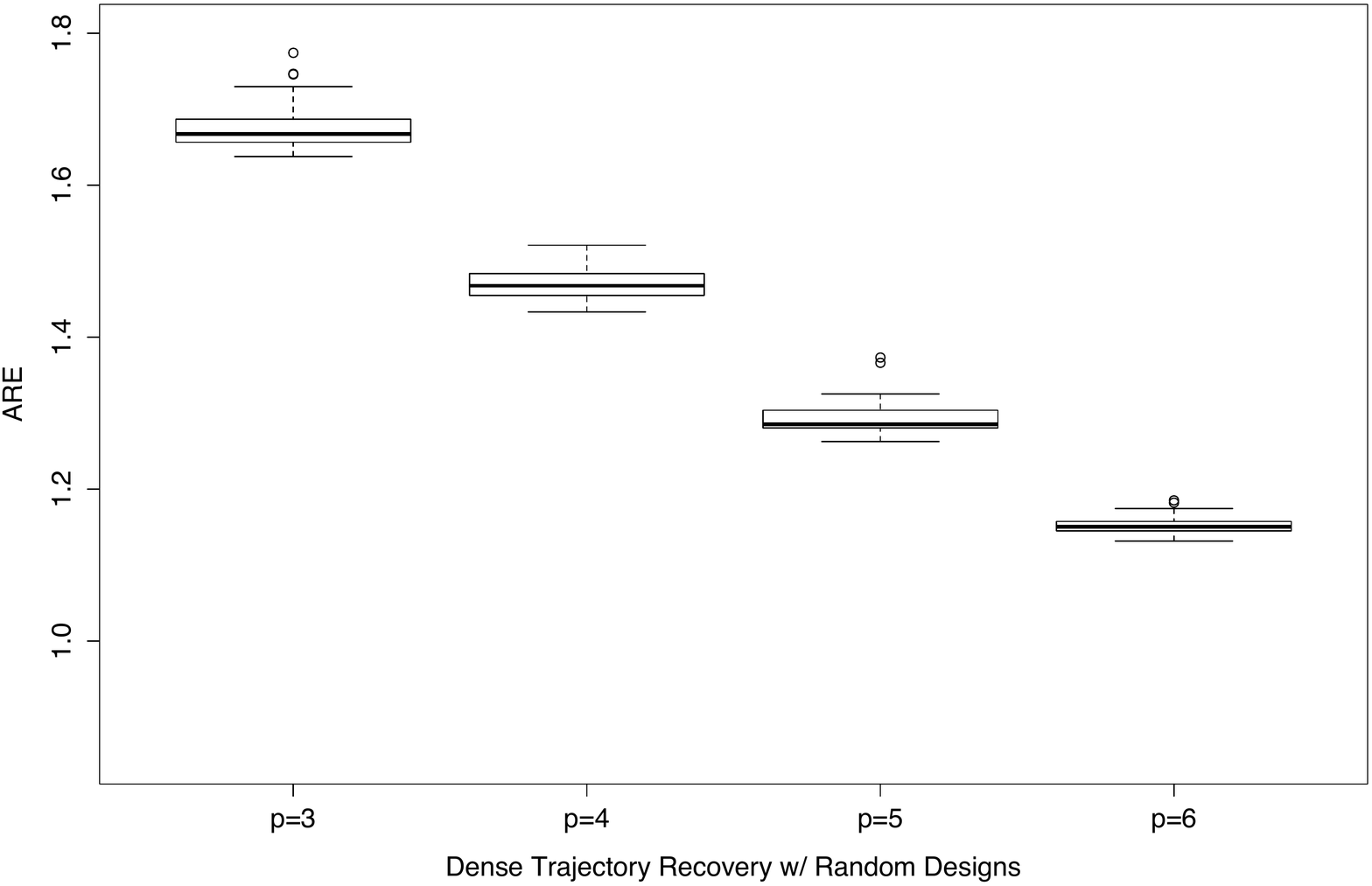}
\includegraphics[scale=0.25]{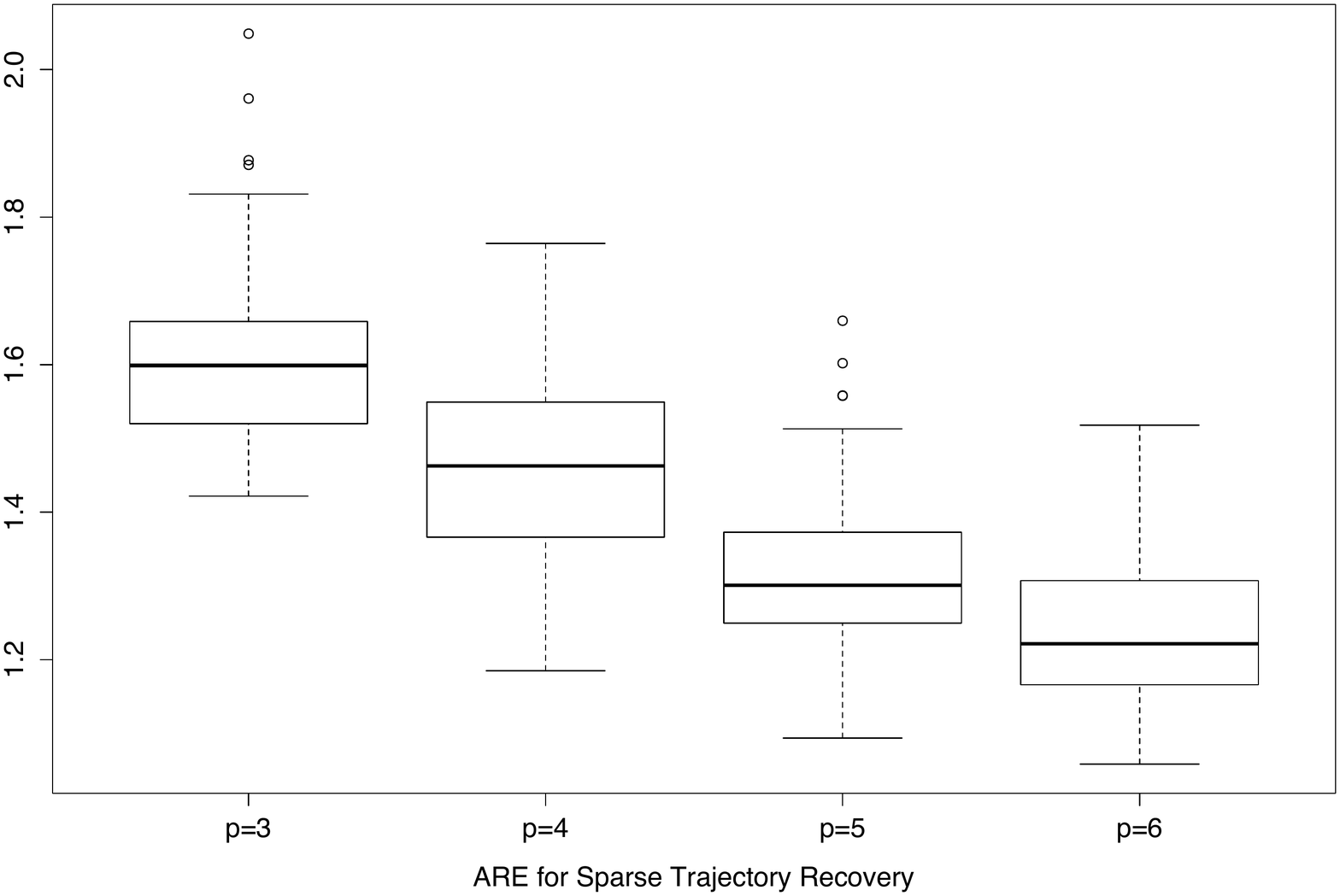}
\includegraphics[scale=0.25]{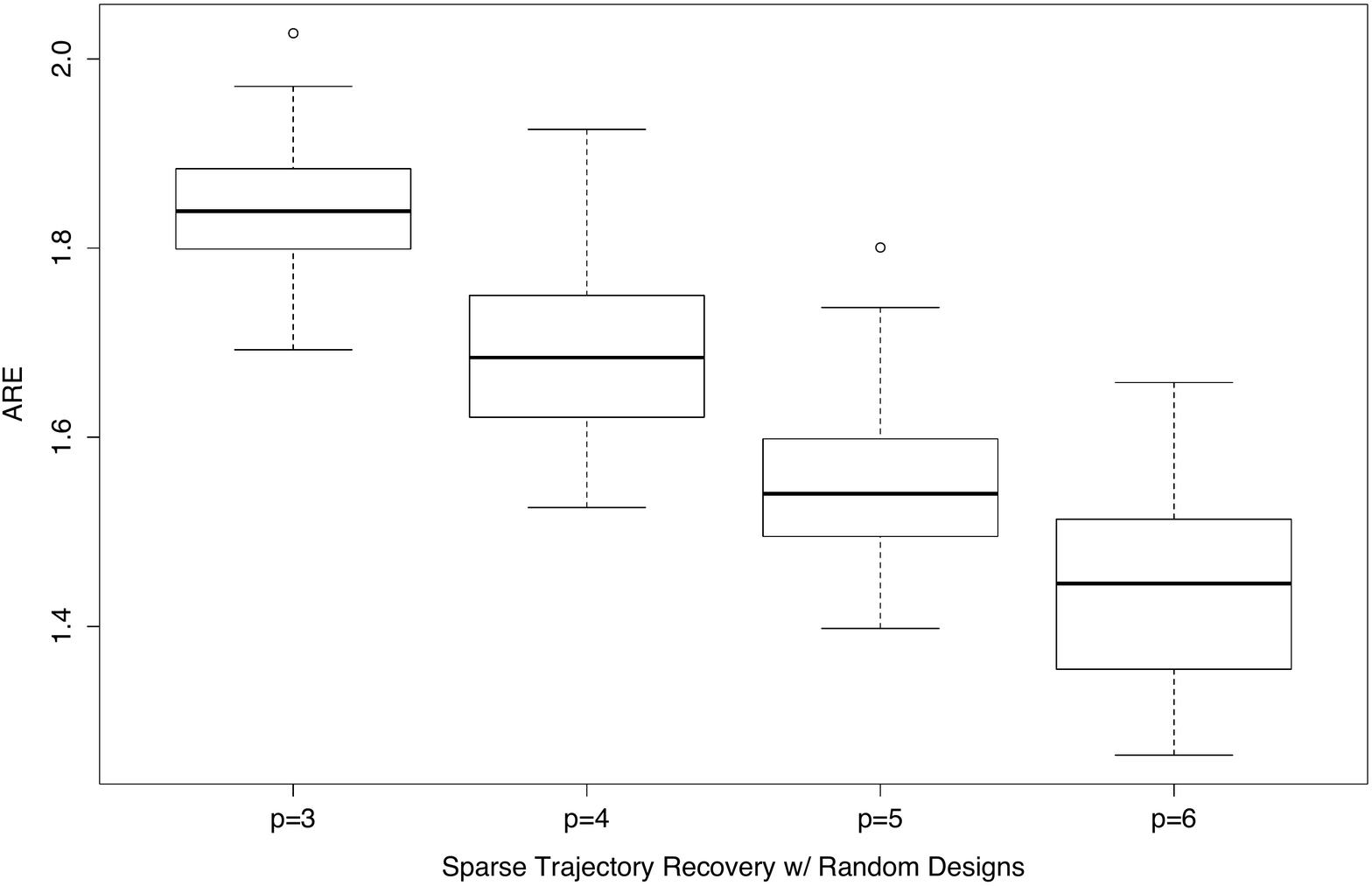}
\caption{Boxplots of ARE from 100 simulation runs for trajectory recovery for both dense and sparse scenarios with sequential search, for $p=3,4,5,6$. The first (second) row corresponds to a dense (sparse) scenario; left (right) column is for the optimal design (median performance random design).}
\end{figure}
\end{center}
\ed


\begin{thebibliography}{28}
\expandafter\ifx\csname natexlab\endcsname\relax\def\natexlab#1{#1}\fi
\expandafter\ifx\csname url\endcsname\relax
  \def\url#1{\texttt{#1}}\fi
\expandafter\ifx\csname urlprefix\endcsname\relax\def\urlprefix{URL }\fi

\bibitem[{Anisimov \textit{et~al.}(2007)Anisimov, Fedorov and Leonov}]{anis:07}
Anisimov, V.~V., Fedorov, V.~V. and Leonov, S.~L. (2007) Optimal design of
  pharmacokinetic studies described by stochastic differential equations.
\newblock In \textit{mODa 8-Advances in Model-Oriented Design and Analysis},
  9--16. Springer.

\bibitem[{Cardot \textit{et~al.}(1999)Cardot, Ferraty and Sarda}]{card:99}
Cardot, H., Ferraty, F. and Sarda, P. (1999) Functional linear model.
\newblock \textit{Statistics and Probability Letters}, \textbf{45}, 11--22.

\bibitem[{Carey \textit{et~al.}(2002)Carey, Liedo, Harshman, Zhang, M{\"u}ller,
  Partridge and Wang}]{Carey:2002}
Carey, J.~R., Liedo, P., Harshman, L., Zhang, Y., M{\"u}ller, H.-G., Partridge,
  L. and Wang, J.-L. (2002) Life history response of mediterranean fruit flies
  to dietary restriction.
\newblock \textit{Aging Cell}, \textbf{1}, 140--148.

\bibitem[{Delaigle \textit{et~al.}(2012)Delaigle, Hall and Bathia}]{hall:12:1}
Delaigle, A., Hall, P. and Bathia, N. (2012) Componentwise classification and
  clustering of functional data.
\newblock \textit{Biometrika}, \textbf{99}, 299--313.

\bibitem[{Fanaee-T and Gama(2013)}]{Fanaee:13}
Fanaee-T, H. and Gama, J. (2013) Event labeling combining ensemble detectors
  and background knowledge.
\newblock \textit{Progress in Artificial Intelligence}, 1--15.

\bibitem[{Fedorov and Leonov(2013)}]{fedo:13}
Fedorov, V.~V. and Leonov, S.~L. (2013) \textit{Optimal design for nonlinear
  response models}.
\newblock CRC Press.

\bibitem[{Ferraty \textit{et~al.}(2010)Ferraty, Hall and Vieu}]{Ferr:10}
Ferraty, F., Hall, P. and Vieu, P. (2010) Most-predictive design points for
  functional data predictors.
\newblock \textit{Biometrika}, \textbf{97}, 807--824.

\bibitem[{Guo(2004)}]{guo:04:1}
Guo, W. (2004) Functional data analysis in longitudinal settings using
  smoothing splines.
\newblock \textit{Statistical Methods in Medical Research}, \textbf{13},
  49--62.

\bibitem[{Hall \textit{et~al.}(2008)Hall, M{\"u}ller and Yao}]{hall:08}
Hall, P., M{\"u}ller, H.-G. and Yao, F. (2008) Modelling sparse generalized
  longitudinal observations with latent gaussian processes.
\newblock \textit{Journal of the Royal Statistical Society: Series B},
  \textbf{70}, 703--723.

\bibitem[{Hall and Vial(2006)}]{hall:06:3}
Hall, P. and Vial, C. (2006) Assessing extrema of empirical principal component
  functions.
\newblock \textit{The Annals of Statistics}, \textbf{34}, 1518--1544.

\bibitem[{He \textit{et~al.}(2000)He, M{\"u}ller and Wang}]{he:2000}
He, G., M{\"u}ller, H. and Wang, J. (2000) Extending correlation and regression
  from multivariate to functional data.
\newblock \textit{Asymptotics in {S}tatistics and {P}robability}, 197--210.

\bibitem[{Hoerl and Kennard(1970)}]{Hoerl:1970}
Hoerl, A.~E. and Kennard, R.~W. (1970) Ridge regression: Biased estimation for
  nonorthogonal problems.
\newblock \textit{Technometrics}, \textbf{12}, 55--67.

\bibitem[{Kirkpatrick \textit{et~al.}(1983)Kirkpatrick, Gelatt and
  Vecchi}]{Kirk:1983}
Kirkpatrick, S., Gelatt, C.~D. and Vecchi, M.~P. (1983) Optimization by
  simulated annealing.
\newblock \textit{Science}, \textbf{220}, 671--680.

\bibitem[{Kouloussis \textit{et~al.}(2011)Kouloussis, Papadopoulos,
  Katsoyannos, M{\"u}ller, Wang, Su, Molleman and
  Carey}]{kouloussis2011seasonal}
Kouloussis, N.~A., Papadopoulos, N.~T., Katsoyannos, B.~I., M{\"u}ller, H.-G.,
  Wang, J.-L., Su, Y.-R., Molleman, F. and Carey, J.~R. (2011) Seasonal trends
  in ceratitis capitata reproductive potential derived from live-caught females
  in greece.
\newblock \textit{Entomologia {E}xperimentalis et {A}pplicata}, \textbf{140},
  181--188.

\bibitem[{Li and Hsing(2010)}]{li:10}
Li, Y. and Hsing, T. (2010) {Uniform convergence rates for nonparametric
  regression and principal component analysis in functional/longitudinal data}.
\newblock \textit{Annals of Statistics}, \textbf{38}, 3321--3351.

\bibitem[{McKeague and Sen(2010)}]{McKea:2010}
McKeague, I.~W. and Sen, B. (2010) Fractals with point impact in functional
  linear regression.
\newblock \textit{Annals of statistics}, \textbf{38}, 2559.

\bibitem[{Mentre \textit{et~al.}(1997)Mentre, Mallet and Baccar}]{ment:97}
Mentre, F., Mallet, A. and Baccar, D. (1997) Optimal design in random-effects
  regression models.
\newblock \textit{Biometrika}, \textbf{84}, 429--442.

\bibitem[{M\"{u}ller \textit{et~al.}(2001)M\"{u}ller, Carey, Wu, Liedo and
  Vaupel}]{mull:01:3}
M\"{u}ller, H.-G., Carey, J.~R., Wu, D., Liedo, P. and Vaupel, J.~W. (2001)
  Reproductive potential predicts longevity of female {M}editerranean fruit
  flies.
\newblock \textit{Proceedings of the Royal Society B}, \textbf{268}, 445--450.

\bibitem[{Pearson \textit{et~al.}(1997)Pearson, Morrell, Brant, Landis and
  Fleg}]{Pearson:1997}
Pearson, J.~D., Morrell, C.~H., Brant, L.~J., Landis, P.~K. and Fleg, J.~L.
  (1997) Age-associated changes in blood pressure in a longitudinal study of
  healthy men and women.
\newblock \textit{The Journals of Gerontology Series A: Biological Sciences and
  Medical Sciences}, \textbf{52}, M177--M183.

\bibitem[{Ramsay and Silverman(2005)}]{ramsayfunctional}
Ramsay, J. and Silverman, B. (2005) \textit{Functional {D}ata {A}nalysis}.
\newblock Springer, New York.

\bibitem[{Rice(2004)}]{rice2004functional}
Rice, J.~A. (2004) Functional and longitudinal data analysis: Perspectives on
  smoothing.
\newblock \textit{Statistica Sinica}, \textbf{14}, 631--648.

\bibitem[{Rice and Wu(2001)}]{rice:2001}
Rice, J.~A. and Wu, C.~O. (2001) Nonparametric mixed effects models for
  unequally sampled noisy curves.
\newblock \textit{Biometrics}, \textbf{57}, 253--259.

\bibitem[{Shock \textit{et~al.}(1984)}]{Shock:1984}
Shock, N.~W. \textit{et~al.} (1984) Normal human aging: The {B}altimore
  {L}ongitudinal {S}tudy of {A}ging.
\newblock \textit{NIH Publication}.

\bibitem[{Staniswalis and Lee(1998)}]{staniswalis:1998}
Staniswalis, J.~G. and Lee, J.~J. (1998) Nonparametric regression analysis of
  longitudinal data.
\newblock \textit{Journal of the American Statistical Association},
  \textbf{93}, 1403--1418.

\bibitem[{Xiang \textit{et~al.}(2013)Xiang, Qiu and Pu}]{xian:13}
Xiang, D., Qiu, P. and Pu, X. (2013) Nonparametric regression analysis of
  multivariate longitudinal data.
\newblock \textit{Statistica Sinica}, \textbf{23}, 769--789.

\bibitem[{Yao \textit{et~al.}(2005{\natexlab{a}})Yao, M{\"u}ller and
  Wang}]{yao:05:1}
Yao, F., M{\"u}ller, H.-G. and Wang, J.-L. (2005{\natexlab{a}}) Functional data
  analysis for sparse longitudinal data.
\newblock \textit{Journal of the American Statistical Association},
  \textbf{100}, 577--590.

\bibitem[{Yao \textit{et~al.}(2005{\natexlab{b}})Yao, M{\"u}ller and
  Wang}]{yao:05:2}
--- (2005{\natexlab{b}}) Functional linear regression analysis for longitudinal
  data.
\newblock \textit{Annals of Statistics}, \textbf{33}, 2873--2903.

\bibitem[{Zagoraiou and Baldi~Antognini(2009)}]{zago:09}
Zagoraiou, M. and Baldi~Antognini, A. (2009) Optimal designs for parameter
  estimation of the {O}rnstein--{U}hlenbeck process.
\newblock \textit{Applied Stochastic Models in Business and Industry},
  \textbf{25}, 583--600.

\end{thebibliography}

\end{document}